\documentclass[a4,11pt]{article}
\usepackage{graphicx} 
\usepackage{float}
\usepackage{amsmath,amsthm}
\usepackage{amsfonts}
\usepackage{latexsym}
\usepackage{amssymb}
\usepackage{setspace}
\usepackage{comment}
\usepackage{cases}
\usepackage{color}
\makeatletter
 
  \@addtoreset{equation}{section}
 \makeatother

\begin{document}
\title{A Mean Field Game Approach to Equilibrium Pricing\\
with Market Clearing Condition~\footnote{
Forthcoming in {\it SIAM Journal on Control and Optimization}.
All the contents expressed in this research are solely those of the author and do not represent any views or 
opinions of any institutions. The author is not responsible or liable in any manner for any losses and/or damages caused by the use of any contents in this research.
}
}

\author{Masaaki Fujii\footnote{Quantitative Finance Course, Graduate School of Economics, The University of Tokyo. }, \quad
Akihiko Takahashi\footnote{Quantitative Finance Course, Graduate School of Economics, The University of Tokyo. }
}
%\begin{center}
\date{ 
First version:  6 March, 2020\\
This version:   25 September, 2021
}
%\end{center}
\maketitle

%%%%%%    TEXT START    %%%%%%

%%%%%%      Macros      %%%%%%
%nakamacro.tex(H120522;0730)
%\documentstyle[11pt]{article}
%\setlength{\textwidth}{10.5in}
%\setlength{\oddsidemargin}{0in}
%\setlength{\topmargin}{-0.52in}
%\setlength{\textheight}{9.0in}
%\setlength{\footskip}{0.7in}

\newtheorem{definition}{Definition}[section]
\newtheorem{assumption}{Assumption}[section]
\newtheorem{condition}{$[$ C}
\newtheorem{lemma}{Lemma}[section]
\newtheorem{proposition}{Proposition}[section]
\newtheorem{theorem}{Theorem}[section]
\newtheorem{remark}{Remark}[section]
\newtheorem{example}{Example}[section]
\newtheorem{corollary}{Corollary}[section]
%--------------------------------------------------------------------------
%BOLD FACES
\def\n{{\bf n}}
\def\A{{\bf A}}
\def\B{{\bf B}}
\def\C{{\bf C}}
\def\D{{\bf D}}
\def\E{{\bf E}}
\def\F{{\bf F}}
\def\G{{\bf G}}
\def\H{{\bf H}}
\def\I{{\bf I}}
\def\J{{\bf J}}
\def\K{{\bf K}}
\def\L{{\bf L}}
\def\M{{\bf M}}
\def\N{{\bf N}}
\def\O{{\bf O}}
\def\P{{\bf P}}
\def\Q{{\bf Q}}
\def\R{{\bf R}}
\def\S{{\bf S}}
\def\T{{\bf T}}
\def\U{{\bf U}}
\def\V{{\bf V}}
\def\W{{\bf W}}
\def\X{{\bf X}}
\def\Y{{\bf Y}}
\def\Z{{\bf Z}}
\def\cala{{\cal A}}
\def\calb{{\cal B}}
\def\calc{{\cal C}}
\def\cald{{\cal D}}
\def\cale{{\cal E}}
\def\calf{{\cal F}}
\def\calg{{\cal G}}
\def\calh{{\cal H}}
\def\cali{{\cal I}}
\def\calj{{\cal J}}
\def\calk{{\cal K}}
\def\call{{\cal L}}
\def\calm{{\cal M}}
\def\caln{{\cal N}}
\def\calo{{\cal O}}
\def\calp{{\cal P}}
\def\calq{{\cal Q}}
\def\calr{{\cal R}}
\def\cals{{\cal S}}
\def\calt{{\cal T}}
\def\calu{{\cal U}}
\def\calv{{\cal V}}
\def\calw{{\cal W}}
\def\calx{{\cal X}}
\def\caly{{\cal Y}}
\def\calz{{\cal Z}}
%
%YOKUTUKAUMONO
\def\sskip{\hspace{0.5cm}}
\def\simleq{ \raisebox{-.7ex}{\em $\stackrel{{\textstyle <}}{\sim}$} }
\def\leqsim{ \raisebox{-.7ex}{\em $\stackrel{{\textstyle <}}{\sim}$} }
\def\ep{\epsilon}
\def\half{\frac{1}{2}}
\def\iku{\rightarrow}
\def\Iku{\Rightarrow}
\def\ikup{\rightarrow^{p}}
\def\inclusion{\hookrightarrow}
\def\cadlag{c\`adl\`ag\ }
\def\up{\uparrow}
\def\down{\downarrow}
\def\doti{\Leftrightarrow}
\def\douti{\Leftrightarrow}
\def\dochi{\Leftrightarrow}
\def\douchi{\Leftrightarrow}%
%KAIGYOU,ARRAY
\def\yy{\\ && \nonumber \\}
\def\y{\vspace*{3mm}\\}
\def\nn{\nonumber}
\def\be{\begin{equation}}
\def\ee{\end{equation}}
\def\bea{\begin{eqnarray}}
\def\eea{\end{eqnarray}}
\def\beas{\begin{eqnarray*}}
\def\eeas{\end{eqnarray*}}
%
%KONO RONBUN DE TUKAU MONO
\def\hd{\hat{D}}
\def\hv{\hat{V}}
\def\hsd{{\hat{d}}}
\def\hx{\hat{X}}
\def\hsx{\hat{x}}
\def\bsx{\bar{x}}
\def\bsd{{\bar{d}}}
\def\bx{\bar{X}}
\def\ba{\bar{A}}
\def\bb{\bar{B}}
\def\bc{\bar{C}}
\def\bv{\bar{V}}
\def\balpha{\bar{\alpha}}
\def\bbalpha{\bar{\bar{\alpha}}}
\def\combi{\l(\begin{array}{c}\alpha\\ \beta \end{array}\r)}
\def\f{^{(1)}}
\def\s{^{(2)}}
\def\ss{^{(2)*}}
\def\l{\left}
\def\r{\right}
\def\a{\alpha}
\def\b{\beta}
\def\L{\Lambda}
%上に定義されたコマンドは数式モ−ドで用いる。
%--------------------------------------------------

%\def\mg{\mathfrak}

\def\calf{{\cal F}}
\def\wt{\widetilde}
\def\mbb{\mathbb}
\def\ol{\overline}
\def\ul{\underline}
\def\sign{{\rm{sign}}}
\def\wh{\widehat}
\def\vr{\varrho}
\def\p{\prime}
\def\pp{^\prime}
\def\ep{\epsilon}
\def\vep{\varepsilon}
\def\del{\delta}
\def\Del{\Delta}
\def\enu{\mathfrak{n}}
\def\chX{\check{X}}
\def\chY{\check{Y}}
\def\chZ{\check{Z}}
\def\chx{\check{x}}
\def\chy{\check{y}}
\def\chz{\check{z}}
\def\ak{\alpha^{(k)}}
\def\as{\alpha^{(s)}}
\def\Ito{It\^o}

\def\hatl{\widehat{\lambda}}
\def\part{\partial}
\def\ubar{\underbar}
\def\ul{\underline}
\def\ol{\overline}
\def\ha{\widehat{\alpha}}
\def\hc{\widehat{c}}
\def\vp{\varpi}
\def\nn{\nonumber}
\def\be{\begin{equation}}
\def\ee{\end{equation}}
\def\bea{\begin{eqnarray}}
\def\eea{\end{eqnarray}}
\def\beas{\begin{eqnarray*}}
\def\eeas{\end{eqnarray*}}
\def\tbf{\textbf}
\def\bg{\boldsymbol}

\def\bull{$\bullet~$}

\newcommand{\Slash}[1]{{\ooalign{\hfil/\hfil\crcr$#1$}}}
\vspace{-5mm}
%%%%%%%%%%%%%%%%%%%%%%%%%%%%%%
\begin{abstract}
%%%%%%%%%%%%%%%%%%%%%%%%%%%%%% 
In this work, we study an equilibrium-based continuous asset pricing problem which seeks to form a price process endogenously
by requiring it to balance the flow of sale-and-purchase orders in the exchange market,  
where a large number of agents $1\leq i\leq N$ are interacting through the market price.
Adopting a mean field game (MFG) approach, we find a special form of forward-backward stochastic differential equations 
of McKean-Vlasov type with common noise whose solution provides an approximate of the market price.
We show the convergence of the net order flow to zero in the large $N$-limit and get the order of convergence in $N$
under some conditions. 
An extension of the model to a setup with multiple populations, where the agents within each population
share the same cost and coefficient functions but they can be different population by population, is also discussed.
 
%%%%%%%%%%%%%%%%%%%%%%%%%%%%%%%
\end{abstract}
\vspace{0mm}
%%%%%%%%%%%%%%%%%%%%%%%%%%%%%%%%%$

%%%%%%%%%%%%%%%%%%%%%%%%%%%%%%%%%
\section{Introduction}
%%%%%%%%%%%%%%%%%%%%%%%%%%%%%%%%%
One of the most important problems in financial economics is to understand how the 
asset price processes are formed through the interaction among a large number of 
rational competitive agents.
In this paper, using a stylized model of security exchange,  we try to explicitly form an 
approximate market price process which balances the flow of sale-and-purchase orders 
from a large number of  rational financial firms. If we directly force the price process 
to balance the net order flow, the strategies of the agents become strongly coupled 
and the problem is hardly solvable. In fact,  it is even unclear how to make the 
cost functions of the agents well-defined,  since the market price results in a very complicated
recursive functional of strategies of all the agents that makes it difficult to guarantee
the convexity of the cost functions. In order to circumvent this problem, we make use of the 
recent developments of mean field games.

Since its inception brought by the pioneering works of Lasry \& Lions~\cite{Lions-1,Lions-2, Lions-3} and Huang, Malhame \& Caines~\cite{Caines-Huang}, mean field game has rapidly developed into one of the most actively studied topics in 
the field of probability theory, applied mathematics, engineering,  finance and economics.
The greatest strength of the mean field game approach is to render notoriously difficult problems
of stochastic differential games among many agents tractable by transforming it to a simpler form 
of stochastic control problems. There exist two approaches to the mean field games, one 
is analytic approach using partial differential equations (PDEs), and the others is probabilistic approach 
based on forward-backward stochastic differential equations (FBSDEs).
For details of analytic approach and its applications, the interested readers may consult the works 
of Bensousssan, Frehse \& Yam~\cite{Bensoussan-mono}, Gomes, Nurbekyan \& Pimentel~\cite{Gomes-eco},
Achdou et.al.~\cite{Achdou-Lions}, Gomes, Pimentel \& Voskanyan~\cite{Gomes-reg} and also Kolokoltsov \& Malafeyev~\cite{Kolokoltsov-mono}.
On the other hand, the probabilistic approach was developed by the series of works of Carmona \& Delarue~\cite{Carmona-Delarue-MKV, 
Carmona-Delarue-MFG, Carmona-Delarue-MFTC} and the recent two volumes of monograph~\cite{Carmona-Delarue-1,
Carmona-Delarue-2} provide its full mathematical details and many references for a wide array of 
applications of mean field games.

There are many macroeconomic applications of mean field games, in particular, those focusing on
general equilibrium models on  growth, inequality and unemployment, etc. See \cite{BM-2,  Aiyagari, Bayraktar, BM-1, Krusell-Smith} for examples.
On the other hand, in the current paper, we are interested in equilibrium price formation in the financial market 
i.e.~the pricing of securities endogenously using a model of rational financial firms (agents) under the market clearing condition.
Interestingly, from the perspective of equilibrium price formation, the number of applications of mean field games is rather limited.
In most of the existing literature, authors  have given a response function of the price process exogenously and searched
an approximate Nash equilibrium among the agents. See, for example, applications to  optimal trading as well as liquidation of portfolio,
exploitation of exhaustible resources and related issues among many agents~\cite{ Fu-Horst-1, Fu-Horst-2, Fu, Lehalle}, or applications 
to electricity pricing with smart grids~\cite{Matoussi, Djehiche-E}.  
Another approach is to assume demand or supply function exogenously without directly solving the 
individual agent optimization problem. For example, in the work \cite{Gueant-Oil},
the authors treat explicitly the balance of demand and supply in the oil market, but the demand is exogenously given 
as a function of the oil price. 
Although these approaches  make the setup nicely fit to the concept of  Nash equilibrium,
which is the standard solution concept in the mean field games,  the market clearing
equilibrium cannot be investigated anymore.  In fact, the endogenous relation between the price processes
and the properties of individual agent is left unknown. 
One notable exception is the work of Gomes \& Saude~\cite{Gomes-Saude}, 
in which the authors explicitly force demand and supply to balance and endogenously construct the market clearing 
electricity price.  They use the analytic approach and the resultant equilibrium price process 
becomes deterministic due to the absence of common noise.

In the current paper, we extend the work  \cite{Gomes-Saude} by adopting the probabilistic approach.
In order to understand the price processes, in particular those of financial assets, including systemic signals
which impact all the agents is crucially important. 
We find an interesting form of FBSDEs of McKean-Vlasov type with common  noise as a limit problem. 
Although it involves dependence in conditional law, it only appears as a conditional expectation.
This allows us to adopt the well-known Peng-Wu's continuation method~\cite{Peng-Wu} to prove the 
existence of a unique strong solution. The resultant candidate of the market price process
is derived completely endogenously by the optimal trading strategies of the agents 
facing systemic information (including 
securities' coupon stream) as well as idiosyncratic noise.
Another benefit of probabilistic approach is that it  allows us to  quantify
the relation between the actual game with finite number of agents and its large population limit.
In a similar manner to the standard mean field games in proving $\varepsilon$-Nash equilibrium, 
we show that the solution of the mean-field limit problem actually provides
asymptotic market clearing in the large-$N$ limit.  Under additional integrability conditions, 
Glivenko-Cantelli convergence theorem in the Wasserstein distance
even provides a specific order of convergence in terms of the number of agents $N$.
It is also possible to extend the model to the situation with multiple populations
where the agents share the same cost and coefficient functions within each population 
but they can be different population by population. This will provide an important tool
to study the price formation in the presence of different type of agents such as Buy-side 
and Sell-side institutions, for example. 
%Due to the page limit for the journal,
%the details for the multi-population extension are given in the preprint version \cite[Section 6]{Fujii-Takahashi-preprint}.

The organization of the paper is as follows: After explaining the notations in Section~\ref{sec-notation}, 
we give an intuitive derivation of the limit problem from the game of finite number of agents in Section~\ref{sec-intuitive},
which motivates the readers to study the special form of FBSDEs of MKV-type.
The solvability of the FBSDE is studied in Section~\ref{sec-solvability}. Using the derived regularity of the solution, 
we prove the asymptotic market clearing in Section~\ref{sec-asymptotic}.
In Section~\ref{sec-multiple-p}, we discuss the extension of the model to the setup with multiple populations.
Finally, in Section~\ref{sec-conclude}, we give concluding remarks. We discuss further
extensions of the model and future directions of research.

%%%%%%%%%%%%%%%%%%%%%%%%%%%%%%%%%
\section{Notations}
\label{sec-notation}
%%%%%%%%%%%%%%%%%%%%%%%%%%%%%%%%
We introduce (N+1) complete probability spaces:
\bea
(\ol{\Omega}^0, \ol{\calf}^0,\ol{\mbb{P}}^0) \quad {\rm{and}} \quad (\ol{\Omega}^i,\ol{\calf}^i,\ol{\mbb{P}}^i)_{i=1}^N~,\nn
\eea
endowed with filtrations $\ol{\mbb{F}}^i:=(\ol{\calf}_t^i)_{t\geq 0}$, $i\in\{0,\cdots, N\}$.
Here, $\ol{\mbb{F}}^0$ is the completion of the 
filtration generated by $d^0$-dimensional Brownian motion $\bg{W}^0$ (hence right-continuous)
and, for each $i\in\{1,\cdots, N\}$,  $\ol{\mbb{F}}^i$ is the complete and right-continuous augmentation of the filtration
generated by $d$-dimensional Brownian motions $\bg{W}^i$
as well as a $\bg{W}^i$-independent $n$-dimensional  square-integrable random variables $(\xi^i)$. $(\xi^i)_{i=1}^N$
are supposed to have the same law. We also introduce the product probability spaces
\be
\Omega^i=\ol{\Omega}^0\times \ol{\Omega}^i, \quad \calf^i, \quad \mbb{F}^i=(\calf_t^i)_{t\geq 0}, \quad \mbb{P}^i, ~i\in\{1,\cdots, N\}\nn
\ee
where $(\calf^i,\mbb{P}^i)$ is the completion of $(\ol{\calf}^0\otimes \ol{\calf}^i, \ol{\mbb{P}}^0\otimes \ol{\mbb{P}}^i)$
and $\mbb{F}^i$ is the complete and right-continuous augmentation of $(\ol{\calf}_t^0\otimes \ol{\calf}_t^i)_{t\geq 0}$.
In the same way, we  define  the complete probability space $(\Omega,\calf,\mbb{P})$ endowed with $\mbb{F}=(\calf_t)_{t\geq 0}$
satisfying the usual conditions as a product of $(\ol{\Omega}^i,\ol{\calf}^i,\ol{\mbb{P}}^i;\ol{\mbb{F}}^i)_{i=0}^N$. 

Throughout the work, the symbol $L$  denotes a given positive constant, 
the symbol $C$ a general positive constant which may change line by line. 
When we want to emphasize that $C$ depends only on some specific variables, say $a$ and $b$, 
we use the symbol $C(a,b)$. For a given constant $T>0$ and any measurable space $(\Omega,\calg)$
with the filtration $\mbb{G}:=(\calg_t)_{t\geq 0}$, we use the following notations for frequently encountered spaces:\\
\bull $\mbb{L}^2(\calg; \mbb{R}^d)$ denotes the set of $\mbb{R}^d$-valued $\calg$-measurable 
square integrable random variables.\\
\bull $\mbb{S}^2(\mbb{G};\mbb{R}^d)$ is the set of $\mbb{R}^d$-valued $\mbb{G}$-adapted continuous processes $\bg{X}$ satisfying
\bea
||X||_{\mbb{S}^2}:=\mbb{E}\bigl[\sup_{t\in[0,T]}|X_t|^2\bigr]^\frac{1}{2}<\infty~. \nn
\eea
\bull $\mbb{H}^2(\mbb{G};\mbb{R}^d)$ is the set of $\mbb{R}^d$-valued $\mbb{G}$-progressively measurable processes $\bg{Z}$ satisfying
\bea
||Z||_{\mbb{H}^2}:=\mbb{E}\Bigl[\Bigl(\int_0^T |Z_t|^2 dt\Bigr)\Bigr]^\frac{1}{2}<\infty~. \nn
\eea
\bull $\call(X)$ denotes the law of a random variable $X$.\\
$\bullet~\calp(\mbb{R}^d)$ is the set of probability measures on $(\mbb{R}^d,\calb(\mbb{R}^d))$. \\
$\bullet~\calp_p(\mbb{R}^d)$ with $p\geq 1$ is the subset of $\calp(\mbb{R}^d)$ with finite $p$-th moment i.e.,
the set of $\mu\in \calp(\mbb{R}^d)$ satisfying
\bea
M_p(\mu):=\Bigl(\int_{\mbb{R}^d}|x|^p \mu(dx)\Bigr)^\frac{1}{p}<\infty~.\nn
\eea
We always assign $\calp_p(\mbb{R}^d)$ with $(p\geq 1)$ the $p$-Wasserstein distance $W_p$,
which makes  $\calp_p(\mbb{R}^d)$ a complete separable metric space.
It is defined by, for any $\mu, \nu\in \calp_p(\mbb{R}^d)$, 
\bea
W_p(\mu,\nu):={\inf}_{\pi\in\Pi_p(\mu,\nu)}\Bigl[\Bigl(\int_{\mbb{R}^d\times \mbb{R}^d} |x-y|^p \pi(dx,dy)\Bigr)^\frac{1}{p}\Bigr]
\label{def-W}
\eea
where $\Pi_p(\mu,\nu)$ denotes the set of probability measures in $\calp_p (\mbb{R}^d\times \mbb{R}^d)$
with marginals $\mu$ and $\nu$. For more details,  see \cite[Chapter 5]{Carmona-Delarue-1}. 
We frequently omit the arguments such as $(\calg, \mbb{G},\mbb{R}^d)$ in the above definitions when there is no confusion 
from the context.

%%%%%%%%%%%%%%%%%%%%%%%%%%%%%%%%%
\section{Intuitive Derivation of the Mean Field Problem}
\label{sec-intuitive}
%%%%%%%%%%%%%%%%%%%%%%%%%%%%%%%%
In this section, in order to introduce the special form of forward-backward stochastic differential equations
of McKean-Vlasov type  to be studied in this paper, 
we give a heuristic derivation of the mean-field limit problem from the corresponding equilibrium problem with
finite number of agents. 
%%%%%%
\subsection{Description of the problem}
%%%%
As a motivating example, we consider the equilibrium-based pricing problem 
of $n$ types of  securities labeled by $k$,  $1\leq k\leq n$, which are 
continuously traded via the securities exchange in the presence of a large number of homogeneous financial firms (agents)
indexed by $i$,  $1\leq i\leq N$.  
Every agent is supposed to have many small individual clients
who can trade the securities only with the agent via the over-the-counter (OTC) market and have no direct access to the
exchange.\footnote{In fact,  only credit-worthy registered financial firms are allowed to directly participate in the securities exchange.
The individual investors and non-financial firms can trade the securities with these registered firms 
playing the role of financial intermediaries. This is called the over-the-counter (OTC) market.}
We denote the market price process of the $n$ securities by an $\mbb{R}^n$-valued process $(\vp_t)_{t\in[0,T]}$,
the detailed mathematical properties of which are to be discussed later. Here, $(\vp_t)^k$ denotes the market price of the $k$th security at time $t$.
In our model, the state process $(X_t^i)_{t\in[0,T]}$ of each agent $i$,  $1\leq i\leq N$, is given by the time evolution of 
his/her position size in the $n$ securities.  For example, let us suppose that the $k$th security is an equity of a certain company.
Then $(X_t^i)^k$ denotes the number of shares of the equity possessed by the $i$th agent at time $t$.
If it is negative, it means that the agent is taking the {\it short} position.
Each agent $i$, $1\leq i\leq N$,  controls the trading speed of the securities $(\alpha_t^i)_{t\in[0,T]}$ via the  exchange
within some space of admissible strategies $\mbb{A}^i$.
More precisely,  $(\alpha^i_t)^k dt$, $1\leq k\leq n$, denotes the number of shares of the $k$th security
bought (or sold if negative) within the time interval $[t,t+dt]$ by the $i$th agent.
In addition to the trading via the exchange, the position size of each agent is affected 
by his/her market making via the OTC market with individual clients.
Although,   in the real market,  each financial firm dynamically controls bid-offer spreads in order to earn trading fees
and to affect the order flows  from his/her clients in a favorable manner to his/her profit, we treat, in this work, the order flows via the OTC market exogenous
and concentrate on the optimal trading problem via the securities exchange for simplicity.
We denote by $(c_t^0)_{t\geq 0}\in \mbb{H}^2(\ol{\mbb{F}}^0;\mbb{R}^n)$ with $c_T^0\in \mbb{L}^2(\ol{\calf}_T^0;\mbb{R}^n)$
the cash flows from the securities or the market news commonly available to all the agents, 
while by $(c_t^i)_{t\geq 0}\in \mbb{H}^2(\ol{\mbb{F}}^i;\mbb{R}^n)$ with $c_T^i\in \mbb{L}^2(\ol{\calf}_T^i;\mbb{R}^n)$ 
some independent factors and news affecting only 
on the agent $i$.\footnote{The dimensions of $c^0$ and $c^i$ are chosen to be $n$ only for the notational simplicity. One can assign 
any fixed dimensions for them so that they can represent any factors that affect the agents' cost functions. }
Moreover,  we assume that $(c_t^i)_{t\geq 0}$  have the common law for all $1\leq i\leq N$.

Let us introduce the measurable functions, $l:[0,T]\times (\mbb{R}^n)^3\rightarrow \mbb{R}^n$,
$\sigma_0:[0,T]\times (\mbb{R}^n)^3\rightarrow \mbb{R}^{n\times d_0}$ and $\sigma:[0,T]\times (\mbb{R}^n)^3\rightarrow \mbb{R}^{n\times d}$.
Using them, we now express the state dynamics of each agent $i$, $1\leq i\leq N$, by
\bea
dX_t^i=\bigl(\a_t^i+l(t,\vp_t,c_t^0, c_t^i)\bigr)dt+\sigma_0(t,\vp_t,c_t^0, c_t^i)dW_t^0+\sigma(t,\vp_t,c_t^0,c_t^i)dW_t^i \nn
\eea
with $X_0^i=\xi^i \in \mbb{L}^2(\ol{\calf}_0^i;\mbb{R}^n)$. $\xi^i$ denotes the initial position size of the $i$th agent
and is assumed to be independently and identically distributed (i.i.d.) among $1\leq i\leq N$.
In addition to $\alpha_t^i dt$ representing the change due to the direct trading via the exchange,
there also exist contributions from the order flows via the OTC market: $l(t,\vp_t, c_t^0,c_t^i)dt$ and $\bigl(\sigma_0(t,\vp_t,c_t^0,c_t^i)dW_t^0$ 
$,\sigma(t,\vp_t,c_t^0,c_t^i)dW_t^i\bigr)$ denote their finite and infinite variation parts, respectively.
We naturally expect that these order flows are dependent on the price of the securities, common as well as idiosyncratic informations.
Suppose,  for example, $l^k(t,\vp_t,c_t^0,c_t^i)<0$.  This means that the clients of the $i$th agent  
are buying the $k$th security from the agent via the OTC market with the net speed $|l^k(t,\vp_t,c_t^0,c_t^i)|$ at time $t$. 
The two infinite variation terms represent the noise in the order flows.

Under such an environment, each agent tries to minimize his/her cost by controlling the trading speed.
We suppose that the problem for each agent $1\leq i\leq N$ is given by
\be
\inf_{\bg{\a^i}\in \mbb{A}^i}J^i(\bg{\a}^i) 
\label{agent-P}
\ee
with some cost functional 
\bea
J^i(\bg{\a}^i):=\mbb{E}\Bigl[\int_0^T f(t,X_t^i,\a_t^i,\vp_t,c_t^0, c_t^i)dt+g(X_T^i,\vp_T,c_T^0,c_T^i)\Bigr]~. \nn
\eea
The space of admissible strategies $\mbb{A}^i$ of the agent $i$ is assumed to be $\mbb{H}^2\bigl((\sigma\{\vp_s:s\leq t\})\vee \calf_t^i)_{t\geq 0};
\mbb{R}^n\bigr)$ i.e. 
the set of processes $(\a^i_t)_{t\geq 0}$ satisfying
\be
\mbb{E}\int_0^T |\a_t^i|^2 dt<\infty~\nn
\ee
adapted to the filtration generated by the common and his/her idiosyncratic shocks as well as  the market price process 
of the securities. 

\begin{remark}[possible choices of the duration $T$]
Since most of the financial firms set up the policies on profit, risk and storage level of securities
for each fiscal period, a natural choice of the duration  $T$ of the model is (quarter of) a year. 
In fact, financial firms typically reduce the size of securities' position significantly at the end of each fiscal year,  for regulatory as well
as accounting reasons, and announce new budgetary goals to the employees for the next year.
If one is interested in the price dynamics in shorter scale among high-frequency traders, an appropriate choice for  the duration $T$
can be much shorter.  For analyzing the long-term behavior of price dynamics, infinite time horizon may become important~\cite{Bayraktar-2}.
We leave its application to equilibrium price formation for future research.
\end{remark}

We introduce the 
following cost functions; $f:[0,T]\times (\mbb{R}^n)^5\rightarrow \mbb{R}$, $g:(\mbb{R}^n)^4\rightarrow \mbb{R}$,
$\ol{f}:[0,T]\times (\mbb{R}^n)^4\rightarrow \mbb{R}$ and $\ol{g}:(\mbb{R}^n)^3\rightarrow \mbb{R}$, which 
are measurable functions such that
\bea
&&f(t,x,\a,\vp,c^0,c):=\langle \vp,\a\rangle+\frac{1}{2}\langle \a,\L\a\rangle+\ol{f}(t,x,\vp,c^0,c), \nn \\
&&g(x,\vp,c^0,c):=-b \langle \vp,x\rangle+\ol{g}(x,c^0,c)~. \nn
\eea
In the first part of the paper, we assume that these functions are common across all the agents.
In the above expression, $f$ and $g$ denote the running and the terminal costs, respectively.
Let us explain the economic meaning of each term.
By buying (or selling if negative) with speed $\alpha_t$, each agent pays (or receives if negative)
$\langle \alpha_t, \vp_t\rangle dt$ amount of cash in the time interval $[t,t+dt]$.
In addition to this direct cost,  we suppose that each agent has to pay the service fees 
to the securities exchange $\langle \alpha_t, \Lambda \alpha_t\rangle dt$ where
$\Lambda$ is an $n\times n$ positive definite matrix. 
These costs are represented by the first two terms of the function $f$.
In addition, each agent is subject to the costs incurred by the financial risk as well as the appropriate 
inventory management of his/her position.
The cost function $\ol{f}$ is supposed to represent these effects,
which are dependent on the position size, cash flows, prices of the securities
as well as any relevant news available to each agent. In particular,  we can make the agents more risk averse
by assigning stronger penalty on their nonzero position $|X|$ of the securities.
The first term of $g$ denotes
the mark-to-market value at the closing time with some discount factor $b<1$.\footnote{We shall see that the condition $b<1$
is necessary to obtain well-defined terminal condition for the limit problem.}
The cost function $\ol{g}$ puts some penalty on the position size at the terminal time $T$.

If the price process $(\vp_t)_{t\in[0,T]}$ is given exogenously (for example,  by the Black-Scholes model $dS_t=S_t(\mu dt+\sigma dW_t), t\geq 0$), 
the problem $(\ref{agent-P})$ is just the standard optimal portfolio
problem with a stochastic inventory process. In this case, the problem can be considered independently for each agent.
Using the position size as the state process with a linear control term is quite popular in the literature.
For example, one can find similar setups in optimal liquidation problems~\cite{AC, Jeanblanc-Kruse, Schied}
and also in their mean-field extensions~\cite{Fu-Horst-1, Fu-Horst-2, Fu, Lehalle}. By interpreting the variable $X$
as the amount of goods or energy produced and $\alpha$ as the production speed,  similar setups 
for the state process can also be found in papers studying economic problems. See, for example,  \cite{Gueant-Oil, Matoussi} and references therein. 
An important feature in the existing literature is that the optimal control or a Nash equilibrium among 
the agents is searched in the market where the price process or its response function
to the agents' actions is exogenously given.

In contrast, in the current paper, we want to determine the price process itself endogenously by 
the fundamental condition of the  market.
As its name suggests, the total number of shares of the securities being purchased by agents via the securities {\it exchange}
at a given time
must be equal to the number of those being sold by the others via the {\it exchange} at the same time.
This balance between the sales and purchases orders must hold at any point in time.
This is called {\it the market clearing condition} and is expressed by 
\bea
\sum_{i=1}^N \ha_t^i=0~, \qquad dt\otimes d\mbb{P}{\rm -a.e.}
\label{orig-clearing}
\eea
where $\wh{\alpha}^i$ is the optimal control of the $i$th agent.
Our problem is to find an appropriate price process $(\vp_t)_{t\geq 0}$ so that it achieves the market clearing condition $(\ref{orig-clearing})$ among the rational agents. This is the problem of {\it equilibrium price formation}, 
which is one of the central themes in  financial economics.\footnote{ In the standard economic theories, the market clearing 
is used to define the competitive equilibrium not only for securities but also for any goods with appropriate modifications. 
See, for example, \cite[Chapter 10]{M-C}.}

\subsection{Intuitive derivation of the mean-field limit}
Intuitively, by a simple economic argument, it is not hard to imagine that the market price automatically
adjusts to the higher values if the excess demand is positive and does so conversely in the opposite case.
However, understanding the problem in rigorous mathematical sense is not at all easy even though we have already put simplistic conditions 
in which the cost functions as well as the 
coefficient functions are the same across the agents.
Due to the clearing condition $(\ref{orig-clearing})$, we cannot adopt the open-loop equilibrium approach.
This means, in particular, if we assume the presence of some feedback effects, 
then the price process $(\vp_t)_{t\geq 0}$ becomes a complicated functional of the agents' trading strategies
and hence the problem for each agent is highly recursive with respect to $(\a^i_t)_{t\geq 0, 1\leq i\leq N}$. 
It is even unclear how to guarantee the cost function well-defined by making it convex  with respect to the controls.

In order to obtain some insight, let us consider a much simpler situation.
It is natural to suppose that the impact to the market price process $(\vp_t)_{t\in[0,T]}$ from the trading of  each agent
is negligibly small when $N$ is sufficiently large. If an agent considers that his/her market share
is negligibly small, he/she behaves as a {\it price taker}, i.e. considers that the market price process is
not affected by his/her actions.  Let us suppose that this is the case for every agent.
Moreover, $(\vp_t)_{t\in[0,T]}$ is likely to be
given by $\ol{\mbb{F}}^0$-progressively measurable process since the effects from the idiosyncratic parts
from many agents are expected to be canceled out. If this is the case, the problem for each agent $i$ reduces to the 
standard  optimal portfolio problem with a stochastic inventory process of the securities in a given  random 
environment $(\vp_t,c_t^0,c_t^i)_{t\in[0,T]}$.
Note that if $(\vp_t)_{t\in[0,T]}$ is $\ol{\mbb{F}}^0$-adapted, then the space of admissible strategies $\mbb{A}^i$ of the agent $i$, $1\leq i\leq N$, 
becomes $\mbb{H}^2(\mbb{F}^i;\mbb{R}^n)$.
Let us first investigate this simple problem in details.   
We introduce the following conditions.

\begin{assumption}\label{MFG-a}{~}\\ 
%{\rm{(MFG-a)}} 
{\rm (i)} $\L$ is a positive definite $n\times n$ symmetric matrix with $\ul{\lambda}I_{n\times n}\leq \L \leq \ol{\lambda}I_{n\times n}$
in the sense of 2nd-order form where $\ul{\lambda}$ and $\ol{\lambda}$ are some constants
satisfying $0< \ul{\lambda}\leq \ol{\lambda}$.  \\
{\rm (ii)} For any $(t,x,\vp,c^0,c)$, 
\bea
|\ol{f}(t,x,\vp,c^0,c)|+|\ol{g}(x,c^0,c)|\leq L(1+|x|^2+|\vp|^2+|c^0|^2+|c|^2)~.\nn
\eea
{\rm (iii)} $\ol{f}$ and $\ol{g}$ are continuously differentiable in $x$ and satisfy,  for any $(t,x,x\pp, \vp,c^0,c)$,
\be
|\part_x \ol{f}(t,x\pp,\vp,c^0,c)-\part_x\ol{f}(t,x,\vp,c^0,c)|+|\part_x \ol{g}(x\pp,c^0,c)-\part_x \ol{g}(x,c^0,c)|\leq L |x\pp-x|~, \nn
\ee 
and $|\part_x \ol{f}(t,x,\vp,c^0,c)|+|\part_x \ol{g}(x,c^0,c)|\leq L(1+|x|+|\vp|+|c^0|+|c|)$. \\
%(iv) For any $(t,x,\vp,c^0,c)$,
%\bea
%|\ol{f}(t,x,\vp,c^0,c)|+|\ol{g}(x,c^0,c)|\leq L(1+|x|^2+|\vp|^2+|c^0|^2+|c|^2)~.\nn
%\eea
{\rm (iv)} The functions $\ol{f}$ and $\ol{g}$ are convex in $x$ in the sense that for any $(t,x,x\pp,\vp,c^0,c)$,
\bea
&&\ol{f}(t,x\pp,\vp,c^0,c)-\ol{f}(t,x,\vp,c^0,c)-\langle x\pp-x, \part_x \ol{f}(t,x,\vp,c^0,c)\rangle\geq \frac{\gamma^f}{2}|x\pp-x|^2~, \nn \\
&&\ol{g}(x\pp,c^0,c)-\ol{g}(x,c^0,c)-\langle x\pp-x,\part_x \ol{g}(x,c^0,c)\rangle \geq \frac{\gamma^g}{2}|x\pp-x|^2~, \nn
\eea
with some constants $\gamma^f,\gamma^g\geq 0$. \\
{\rm (v)} $l,\sigma_0$ and $\sigma$ satisfy the linear growth condition:
\bea
|(l,\sigma_0,\sigma)(t,\vp,c^0,c)|\leq L(1+|\vp|+|c^0|+|c|) \nn
\eea
for any $(t,\vp,c^0,c)$. \\
{\rm (vi)} $b \in [0,1)$ is a given constant.
\end{assumption}

\begin{remark}
\label{remark-3-1}
If $c^0$ denotes a coupon stream of the securities, one may use  for example, 
\bea
\ol{f}(t,x,\vp,c^0,c)=-\langle c^0,x\rangle+\ol{f}^\prime(t,x,\vp,c)\nn
\eea
as a running cost with an appropriate measurable function $\ol{f}^\prime$.
As for securities with a given maturity $T$ with exogenously specified payoff $c^0$, such as bonds and futures, it is  natural to 
consider
\bea
g(x,c^0)=\ol{g}(x,c^0)=-\langle c^0,x\rangle \nn
\eea
as the terminal cost. Since the securities cease to exist at time $T$ after paying $c^0$, there is no
reason to put the penalty on the terminal position size anymore.
\end{remark}

%The first term $\langle \vp,\a\rangle$ of $f$ denotes the direct cost incurred by the sales and purchase of the securities
%and the second term $\frac{1}{2}\langle\a, \L\a\rangle$ is  some fee to be paid to the exchange depending on the trading speed, or 
%may be interpreted as  some internal cost. The first term of $g$ denotes
%the mark-to-market value at the closing time with some discount factor $b<1$.\footnote{We shall see that the condition $b<1$
%is necessary to obtain well-defined terminal condition for the limit problem.}
%$\ol{f}$ and $\ol{g}$ denote the running as well as the terminal cost which are
%affected by the market price, coupon streams, or the news.

For this problem, the (reduced) Hamiltonian is given by
\bea
H(t,x,y,\a,\vp,c^0,c)=\langle y, \a+l(t,\vp,c^0,c)\rangle+f(t,x,\a,\vp,c^0,c)~.\nn
\eea
Since $\part_\a H(t,x,y,\a,\vp,c^0,c)=y+\vp+\Lambda \a$, the minimizer of the Hamiltonian is
\bea
\ha(y,\vp):=-\ol{\L}(y+\vp)
\label{def-ha}
\eea
where $\ol{\L}:=\L^{-1}$. The adjoint FBSDE associated with the stochastic maximal principle for each agent $1\leq i\leq N$ is thus given by,
\bea
&&dX_t^i=\Bigl(\ha(Y_t^i,\vp_t)+l(t,\vp_t,c_t^0, c_t^i)\Bigr)dt+\sigma_0(t,\vp_t,c_t^0, c_t^i)dW_t^0+\sigma(t,\vp_t,c_t^0,c_t^i)dW_t^i~, \nn \\
&&dY_t^i=-\part_x \ol{f}(t,X_t^i,\vp_t,c_t^0,c_t^i)dt+Z_t^{i,0}dW_t^0+Z_t^{i}dW_t^i~,
\label{agent-FBSDE}
\eea
with $X_0^i=\xi^i$ and $Y_T^i=\part_x g(X_T^i, \vp_T, c_T^0, c_T^i)$.

\begin{theorem}
\label{th-individual}
Under Assumption~\ref{MFG-a} and a given $(\vp_t)_{t\in[0,T]}\in \mbb{H}^2(\ol{\mbb{F}}^0;\mbb{R}^n)$ with 
$\vp_T\in \mbb{L}^2(\ol{\calf}_T^0;\mbb{R}^n)$,
the problem $(\ref{agent-P})$ for each agent is uniquely characterized by the FBSDE $(\ref{agent-FBSDE})$
which is strongly solvable with a unique solution $(X^i,Y^i,Z^{i,0}, Z^{i})\in \mbb{S}^2(\mbb{F}^i;\mbb{R}^n)\times 
\mbb{S}^2(\mbb{F}^i;\mbb{R}^n)\times \mbb{H}^2(\mbb{F}^i;\mbb{R}^{n\times d^0})\times \mbb{H}^2(\mbb{F}^i;\mbb{R}^{n\times d})$.
\begin{proof}
Since the cost functions are jointly convex with $(x,\a)$ and strictly convex in $\a$, the problem 
is the special situation investigated in \cite[Section 1.4.4]{Carmona-Delarue-2}. Note that, in our case, 
the diffusion terms $\sigma_0,\sigma$ are independent of  $(X^i,\a^i)$. Hence the proof is the direct result of \cite[Theorem 1.60]{Carmona-Delarue-2}. 
\end{proof}
\end{theorem}
Using the above solution, the optimal strategy of each agent is given by
\bea
\ha^i_t=-\ol{\L}(Y_t^i+\vp_t),\quad t\in[0,T]~.\nn
\eea
Let us check the market clearing condition. In the current situation, $(\ref{orig-clearing})$ is equivalent to
\bea
\vp_t=-\frac{1}{N}\sum_{i=1}^N Y_t^i\nn
\eea
which is of course inconsistent with the our simplifying assumption that requires $(\vp_t)_{t\geq 0}$ to be an  $\ol{\mbb{F}}^0$-adapted process.
However, in the current setup, for any $t\in[0,T]$, $(Y^i_t)_{i=1}^N$ are exchangeable random variables
due to the construction of the probability space, common coefficient functions, and the fact that $(\xi^i)_{i=1}^N$
as well as $(c_t^i, t\in[0,T])_{i=1}^N$ are assumed to be i.i.d.  Thus De Finetti's theory of exchangeable sequence of random variables  tells
\bea
\lim_{N\rightarrow \infty}\frac{1}{N}\sum_{i=1}^N Y_t^i=\mbb{E}\Bigl[Y_t^1|\bigcap_{k\geq 1}\sigma\{Y_t^j,j\geq k\}\Bigr]\quad{\rm a.s.}\nn
\eea
See for example \cite[Theorem 2.1]{Carmona-Delarue-2}. It also seems natural to expect that the tail $\sigma$-field is
reduced to $\ol{\calf}_t^0$. Therefore we can expect that,  in the large-$N$ limit, the market price of the securities 
may be given by  $\vp_t=-\mbb{E}[Y_t^1|\ol{\calf}_t^0]$.

The above observation motivates us to consider the following FBSDE:
\bea
&&dX_t=\Bigl(\ha\bigl(Y_t,-\mbb{E}[Y_t|\ol{\calf}_t^0]\bigr)+l\bigl(t,-\mbb{E}[Y_t|\ol{\calf}_t^0],c_t^0, c_t\bigr)\Bigr)dt\nn \\
&&\qquad\quad +\sigma_0\bigl(t,-\mbb{E}[Y_t|\ol{\calf}_t^0],c_t^0, c_t\bigr)dW_t^0+\sigma\bigl(t,-\mbb{E}[Y_t|\ol{\calf}_t^0],c_t^0,c_t\bigr)dW_t^1~, \nn \\
&&dY_t=-\part_x \ol{f}\bigl(t,X_t,-\mbb{E}[Y_t|\ol{\calf}_t^0],c_t^0,c_t\bigr)dt+Z_t^{0}dW_t^0+Z_tdW_t^1~, \nn
\eea
with $X_0=\xi$ and 
\be 
Y_T=\frac{b}{1-b}\mbb{E}\bigl[\part_x \ol{g}(X_T,c_T^0,c_T)|\ol{\calf}_T^0\bigr]+\part_x \ol{g}(X_T,c_T^0,c_T). \nn
\ee
To simplify the notation,  we have omitted the superscript $1$ from $Y^1$, $X^1$, $\xi^1$ and $c^1$.
Let us remark on the terminal condition.  $Y_T=\part_x g(X_T,-\mbb{E}[Y_T|\ol{\calf}_T^0],c_T^0,c_T)$
is not yet fully specified. Taking the conditional expectation in the both sides gives 
\bea
\mbb{E}[Y_T|\ol{\calf}_T^0]=b \mbb{E}[Y_T|\ol{\calf}_T^0]+\mbb{E}\bigl[\part_x \ol{g}(X_T,c_T^0,c_T)|\ol{\calf}_T^0\bigr]~, \nn
\eea
which implies $\mbb{E}[Y_T|\ol{\calf}_T^0]=\frac{1}{1-b}\mbb{E}\bigl[\part_x \ol{g}(X_T,c_T^0,c_T)|\ol{\calf}_T^0\bigr]$. 
Substituting this expression for $\mbb{E}[Y_T|\ol{\calf}_T^0]$ in $\part_x g$, 
we get the above specification of the terminal condition.

This is the FBSDE we are going to study in the following. It is of McKean-Vlasov type with common noise, 
and similar to the FBSDEs relevant for the extended mean field games.
In the following, we are going to prove the existence of a unique solution to the above FBSDE under
appropriate conditions and then show that $\bigl(-\mbb{E}[Y_t|\ol{\calf}_t^0]\bigr)_{t\in[0,T]}$ actually provides a 
reasonable approximation of the market clearing price process.
In the large-$N$ limit, the market clearing condition we seek is defined by
\be
\lim_{N\rightarrow \infty} \frac{1}{N}\sum_{i=1}^N \wh{\alpha}_t^i=0, \qquad dt\otimes d\mbb{P}{\rm -a.e.}
\label{large-N-clearing}
\ee
Although it is weaker than $(\ref{orig-clearing})$ by the factor $N^{-1}$, 
this is the standard convention of the market clearing condition used in the economics literature dealing with infinite number of agents.
See,  for example, F\"ollmer (1974)~\cite{Follmer} and references therein. 
Note that, if $(\vp_t)_{t\in[0,T]}\in \mbb{H}^2(\ol{\mbb{F}}^0;\mbb{R}^n)$ with $\vp_T\in \mbb{L}^2(\ol{\calf}_T^0;\mbb{R}^n)$ is given arbitrary,
then the quantity in the left-hand side, which is the average of the excess demand,\footnote{It is standard to define
the excess demand by $\sum_{i=1}^N \alpha_t^i$ including the both signs.} has 
a non-zero limit of $\mbb{L}^2$-moment. See also the discussion in Remark~\ref{remark-clearing}.

%%%%%%%%%%%%%%%%%%%%%%%%%%%%%%%%%%%%%%%%%
\section{Solvability of the mean-field FBSDE}
\label{sec-solvability}
%%%%%%%%%%%%%%%%%%%%%%%%%%%%%%%%%%%%%%%%%
We now investigate the solvability of the FBSDE derived in the last section
\bea
&&dX_t=\Bigl(\ha\bigl(Y_t,-\mbb{E}[Y_t|\ol{\calf}_t^0]\bigr)+l\bigl(t,-\mbb{E}[Y_t|\ol{\calf}_t^0],c_t^0, c_t\bigr)\Bigr)dt\nn \\
&&\qquad\quad +\sigma_0\bigl(t,-\mbb{E}[Y_t|\ol{\calf}_t^0],c_t^0, c_t\bigr)dW_t^0+\sigma\bigl(t,-\mbb{E}[Y_t|\ol{\calf}_t^0],c_t^0,c_t\bigr)dW_t^1~, \nn \\
&&dY_t=-\part_x \ol{f}\bigl(t,X_t,-\mbb{E}[Y_t|\ol{\calf}_t^0],c_t^0,c_t\bigr)dt+Z_t^{0}dW_t^0+Z_tdW_t^1~, 
\label{fbsde-single-p}
\eea
with $X_0=\xi$ and 
\be Y_T=\frac{b}{1-b}\mbb{E}\bigl[\part_x \ol{g}(X_T,c_T^0,c_T)|\ol{\calf}_T^0\bigr]+\part_x \ol{g}(X_T,c_T^0,c_T). \nn
\ee
$\ha$ is defined as in $(\ref{def-ha})$.
$(c^0_t)_{t\geq 0}\in \mbb{H}^2(\ol{\mbb{F}}^0;\mbb{R}^n)$ and $(c_t)_{t\geq 0}\in \mbb{H}^2(\ol{\mbb{F}}^1;\mbb{R}^n)$
with square integrable $c_T^0, c_T$ are given as inputs. Let us remind the notation to write $\xi=\xi^1$ and $c=c^1$.

%\begin{remark}
%Although we can work on the smaller space $(\Omega^1,\calf^1,\mbb{P}^1; \mbb{F}^1)$,  we instead use $(\Omega,\calf,\mbb{P};\mbb{F})$ 
%by treating the natural extension of the processes on the latter so that we can avoid to use $\mbb{E}^{\mbb{P}^1}$, $\mbb{P}^1$-a.s. etc.
%\end{remark}
\subsection{Unique existence for small $T$}
\begin{assumption}%{\rm{(MFG-b)}}\\
\label{MFG-b}
For any $(t,x,c^0,c)\in[0,T]\times (\mbb{R}^n)^3$ and any $\vp,\vp\pp \in \mbb{R}^n$, 
the coefficient functions $l,\sigma_0,\sigma$ and $\ol{f}$ satisfy,  with $L$ and another nonnegative constant $L_\vp$,
\be
\begin{split}
&|(l,\sigma_0,\sigma)(t,\vp,c^0,c)-(l,\sigma_0,\sigma)(t,\vp\pp,c^0,c)|\leq L |\vp-\vp\pp|, \\
&|\part_x \ol{f}(t,x,\vp,c^0,c)-\part_x\ol{f}(t,x,\vp\pp,c^0,c)|\leq L_{\vp} |\vp-\vp\pp|~.\nn
\end{split}
\ee
\end{assumption}
Due to the Lipschitz continuity and the absence of $(Z^0,Z)$ in the diffusion coefficients of the forward SDE, 
we have the following short-term existence result.
\begin{theorem}
\label{th-short-T}
Under Assumptions~\ref{MFG-a}and \ref{MFG-b}, there exists some constant $\tau>0$ which depends only on $(L,L_\vp, \ul{\lambda}, b)$
such that for any $T\leq \tau$, there exists a unique strong solution $(X,Y,Z^0,Z)\in \mbb{S}^2(\mbb{F}^1;\mbb{R}^n)
\times \mbb{S}^2(\mbb{F}^1;\mbb{R}^n)\times \mbb{H}^2(\mbb{F}^1;\mbb{R}^{n\times d^0})\times \mbb{H}^2(\mbb{F}^1;\mbb{R}^{n\times d})$
to the FBSDE $(\ref{fbsde-single-p})$.
\begin{proof}
Although there exist terms involving $\mbb{E}[Y_t|\ol{\calf}_t^0]$,  one can adopt the 
standard technique for the Lipschitz FBSDE. See, for example, the proof for  \cite[Theorem 1.45]{Carmona-Delarue-2}.
\end{proof}
\end{theorem}

\subsection{Unique existence for general $T$}
In order to obtain the existence result for general $T$, we are going to apply the technique developed by Peng \& Wu~\cite{Peng-Wu}.
In the case of the standard optimization problem, the joint convexity in the state and control variables combined with 
strict convexity in the control variable are enough to obtain the unique existence. Interestingly however, 
we need a strict convexity also in the state variable $X$ in our problem. As we shall see, this is because the term $-\mbb{E}[Y_t|\ol{\calf}_t^0]$
which appears due to the clearing condition weakens the convexity.

\begin{assumption}
\label{MFG-c1}{~}\\
%(MFG-c1)\\
{\rm (i)} The functions $\sigma_0$ and $\sigma$ are independent of the argument $\vp$. \\
{\rm (ii)} For any $t\in[0,T]$,  any random variables $x,x\pp, c^0, c\in \mbb{L}^2(\calf;\mbb{R}^n)$ and any sub-$\sigma$-field $\calg\subset \calf$, 
the function $l$ satisfies the monotone condition,  with some positive constant $\gamma^l>0$,
\bea
\mbb{E}\Bigl[\langle l(t,\mbb{E}[x|\calg],c^0,c)-l(t,\mbb{E}[x\pp|\calg],c^0,c),x-x\pp\rangle\Bigr]
\geq \gamma^l  \bold{1}_{\{L_{\vp}>0\}} \mbb{E}\bigl[\mbb{E}[x-x\pp|\calg]^2\bigr]~.\nn
\eea
{\rm (iii)} There exists a strictly positive constant $\gamma$ satisfying
$0<\gamma\leq \Bigl(\gamma^f-\frac{L_\vp^2}{4\gamma^l}\Bigr)\wedge \gamma^g$.
Moreover,  for any $x,x\pp, c^0,c\in \mbb{L}^2(\calf;\mbb{R}^n)$ and any sub-$\sigma$-field $\calg\subset \calf$, 
the function $\ol{g}$ satisfies
\bea
\gamma^g\mbb{E}[|x-x\pp|^2]+\frac{b}{1-b}\mbb{E}\Bigl[\langle \mbb{E}\bigl[\part_x\ol{g}(x,c^0,c)-\part_x\ol{g}(x\pp,c^0,c)
|\calg\bigr],x-x\pp\rangle\Bigr]\geq \gamma \mbb{E}[|x-x\pp|^2]~.\nn
\eea
\end{assumption}
\begin{remark}
If $l$ and $\part_x \ol{g}$ have separable forms such as $h(x)+h^c(c^0,c)$
with some functions $h$ and $h^c$, then the conditions (ii) and (iii) are satisfied when the function $h$ is monotone.
Economically speaking, the condition (ii) roughly implies that the demands from the individual OTC clients for 
the securities decrease when  their  market prices rise. 
\end{remark}

The next theorem is the first main existence result.
\begin{theorem}
\label{th-general-T}
Under Assumptions~\ref{MFG-a}, \ref{MFG-b} and \ref{MFG-c1}, there exists a unique strong solution $(X,Y,Z^0,Z)\in \mbb{S}^2(\mbb{F}^1;\mbb{R}^n)
\times \mbb{S}^2(\mbb{F}^1;\mbb{R}^n)\times \mbb{H}^2(\mbb{F}^1;\mbb{R}^{n\times d^0})\times \mbb{H}^2(\mbb{F}^1;\mbb{R}^{n\times d})$
to the FBSDE $(\ref{fbsde-single-p})$.
\begin{proof}
In order to simplify the notation, let us define the functionals $B,F$ and $G$ for any $y,x,c^0,c\in \mbb{L}^2(\calf;\mbb{R}^n)$  by
\bea
&&B(t,y,c^0,c):=\Bigl(-\ol{\L}(y-\mbb{E}[y|\ol{\calf}_t^0])+l(t,-\mbb{E}[y|\ol{\calf}_t^0],c^0,c)\Bigr), \nn \\
&&F(t,x,y,c^0,c):=-\part_x \ol{f}\bigl(t,x,-\mbb{E}[y|\ol{\calf}_t^0],c^0,c\bigr), \nn \\
&&G(x,c^0,c):=\frac{b}{1-b}\mbb{E}\bigl[\part_x\ol{g}(x,c^0,c)|\ol{\calf}_T^0\bigr]+\part_x\ol{g}(x,c^0,c)~.
\label{BG-notation}
\eea
With the convention $\Del y:=y-y^\prime$, $\Del x:=x-x^\prime$, one can easily confirms
\bea
&&\mbb{E}\bigl[\langle B(t,y,c^0,c)-B(t,y\pp,c^0,c),\Del y\rangle\bigr]\leq -\gamma^l \bold{1}_{\{L_\vp>0\}} \mbb{E}\bigl[\mbb{E}[\Del y|\ol{\calf}_t^0]^2\bigr]~, \nn  \\
&&\mbb{E}\bigl[\langle F(t,x,y,c^0,c)-F(t,x\pp,y\pp,c^0,c),\Del x\rangle\bigr]
\leq -\Bigl(\gamma^f-\frac{L_\vp^2}{4\gamma^l}\Bigr)\mbb{E}[|\Del x|^2]+\gamma^l\bold{1}_{\{L_\vp>0\}}\mbb{E}\bigl[\mbb{E}[\Del y|\ol{\calf}_t^0]^2\bigr], \nn \\
&&\mbb{E}\bigl[\langle G(x,c^0,c)-G(x\pp,c^0,c),\Del x\rangle\bigr]\geq \gamma \mbb{E}[|\Del x|^2], 
\label{peng-wu-condition}
\eea
where the first estimate follows from Assumption~\ref{MFG-c1}(ii) and Jensen's inequality, 
the second from Assumptions~\ref{MFG-a}(iv), \ref{MFG-b}
and Cauchy-Schwarz inequality. The third one is the direct consequence of  Assumption~\ref{MFG-c1}(iii).
%\footnote{Under the setup used for Theorem~\ref{th-individual}, the first quantity would be bounded by $-(\ol{\lambda})^{-1}|\Del y|^2$.
%This strict monotone condition is not available due to the presence of $\mbb{E}[\Del y|\ol{\calf}_t^0]$ terms.}

We first make the following hypothesis: there exists some constant $\vr\in[0,1)$ such that,
for any $(I_t^b)_{t\geq 0}$, $(I_t^f)_{t\geq 0}$ in $\mbb{H}^2(\mbb{F}^1;\mbb{R}^n)$ and any $\eta\in \mbb{L}^2(\calf_T^1;\mbb{R}^n)$,
there exists a unique solution $(x^\vr,y^\vr,z^{0,\vr},z^\vr)\in \mbb{S}^2(\mbb{F}^1;\mbb{R}^n)\times \mbb{S}^2(\mbb{F}^1;\mbb{R}^n) 
\times \mbb{H}^2(\mbb{F}^1;\mbb{R}^{n\times d^0})\times \mbb{H}^2(\mbb{F}^1;\mbb{R}^{n\times d})$ to the FBSDE:
\bea
&&dx_t^\vr=\bigl(\vr B(t,y_t^\vr,c_t^0,c_t)+I_t^b\bigr)dt+\sigma_0(t,c_t^0,c_t)dW_t^0+\sigma(t,c_t^0,c_t)dW_t^1~, \nn \\
&&dy_t^\vr=-\bigl((1-\vr)\gamma x_t^\vr-\vr F(t,x_t^\vr,y_t^\vr, c_t^0,c_t)+I_t^f\bigr)dt+z_t^{0,\vr}dW_t^0+z_t^\vr dW_t^1~, 
\label{shifted-0}
\eea
with $x_0^\vr=\xi$ and $y_T^\vr=\vr G(x_T^\vr,c_T^0,c_T)+(1-\vr)x_T^\vr+\eta$.
Note that when $\vr=0$ we have a decoupled set of SDE and BSDE and hence the hypothesis trivially holds.
Our goal is to extend the $\vr$ up to $1$ by following Peng-Wu's continuation method~\cite{Peng-Wu}.
Now, for an arbitrary set of inputs $(x,y,z^0,z)\in \mbb{S}^2(\mbb{F}^1;\mbb{R}^n)^2 \times \mbb{H}^2(\mbb{F}^1;\mbb{R}^{n\times d^0})\times \mbb{H}^2(\mbb{F}^1;\mbb{R}^{n\times d})$
and constant $\zeta\in(0,1)$, consider
\bea
&&dX_t=\bigl[\vr B(t,Y_t,c_t^0,c_t)+\zeta B(t,y_t,c_t^0,c_t)+I_t^b\bigr]dt+\sigma_0(t,c_t^0,c_t)dW_t^0+\sigma(t,c_t^0,c_t)dW_t^1~, \nn  \\
&&dY_t=-\bigl[(1-\vr)\gamma X_t-\vr F(t,X_t,Y_t, c_t^0,c_t)+\zeta(-\gamma x_t-F(t,x_t,y_t,c_t^0,c_t))+I_t^f\bigr]dt \nn\\
&&\qquad\quad+Z_t^{0}dW_t^0+Z_tdW_t^1~, 
\label{shifted-1}
\eea
with $X_0=\xi$ and $Y_T=\vr G(X_T,c_T^0,c_T)+(1-\vr)X_T+\zeta(G(x_T,c_T^0,c_T)-x_T)+\eta$.
The existence of the solution $(X,Y,Z^0,Z)\in \mbb{S}^2\times\mbb{S}^2\times \mbb{H}^2\times \mbb{H}^2$ is guaranteed by the 
previous hypothesis. We are going to prove the map $(x,y,z^0,z)\mapsto (X,Y,Z^0,Z)$ defined above becomes strict contraction
when $\zeta>0$ is chosen small enough.

For two sets of inputs $(x,y,z^0,z)$ and $(x\pp,y\pp,z^{0\p},z\pp)$, let us denote the 
corresponding solutions to $(\ref{shifted-1})$ by $(X,Y,Z^0,Z)$ and $(X\pp,Y\pp,Z^{0\p},Z\pp)$, respectively.
We put $\Del X_t:=X_t-X_t\pp$, $\Del Y_t:=Y_t-Y_t\pp$ and similarly for the others.
Applying \Ito's formula to $\langle \Del X, \Del Y\rangle$ and using the estimates $(\ref{peng-wu-condition})$, we obtain
\bea
&&\mbb{E}\bigl[\langle \Del X_T,\Del Y_T\rangle\bigr]\leq -\gamma\mbb{E}\int_0^T |\Del X_t|^2dt \nn \\
&&\hspace{30mm}+\zeta C\mbb{E}\int_0^T \Bigl[|\Del Y_t|(|\Del y_t|+\mbb{E}[\Del y_t|\ol{\calf}_t^0])+|\Del X_t|
(|\Del x_t|+\mbb{E}[|\Del y_t|\ol{\calf}_t^0])\Bigr]dt~. \nn
\eea
On the  other hand, from the terminal condition on $\Del Y_T$,  we get
\bea
\mbb{E}\bigl[\langle \Del X_T,\Del Y_T\rangle\bigr]
\geq (\vr \gamma+(1-\vr))\mbb{E}[|\Del X_T|^2]-\zeta C\mbb{E}\bigl[|\Del X_T|(|\Del x_T|+\mbb{E}[|\Del x_T||\ol{\calf}_T^0])\bigr]~.\nn
\eea
In both cases, the constant $C$ is $\vr$-independent.
Let us set $\gamma_c:=\min(1,\gamma)>0$. Then one easily confirms $0<\gamma_c\leq \vr\gamma+(1-\vr)$ for 
any $\vr\in[0,1)$. Then the above estimates yield
\bea
\gamma_c \mbb{E}\Bigl[|\Del X_T|^2+\int_0^T|\Del X_t|^2 dt\Bigr]&\leq& \zeta C\mbb {E}\bigl[|\Del X_T|(|\Del x_T|+\mbb{E}[|\Del x_T||\ol{\calf}_T^0])\bigr]
\nn \\
&&\hspace{-30mm}+\zeta C\mbb{E}\int_0^T\Bigl[|\Del Y_t|(|\Del y_t|+\mbb{E}[\Del y_t|\ol{\calf}_t^0])+|\Del X_t|
(|\Del x_t|+\mbb{E}[|\Del y_t|\ol{\calf}_t^0])\Bigr]dt~.\nn
\eea
Using Young's inequality and a new constant $C$, we get
\bea
\mbb{E}[|\Del X_T|^2]+\mbb{E}\int_0^T |\Del X_t|^2 dt \leq \zeta C\mbb{E}\int_0^T\bigl(|\Del Y_t|^2+(|\Del x_t|^2+|\Del y_t|^2)\bigr)dt+\zeta C\mbb{E}[|\Del x_T|^2]~.
\label{eq-pw-1}
\eea
Treating $X,X\pp$ as inputs, the standard estimates for the Lipschitz BSDEs (see, for example, \cite[Theorem 4.2.3]{Zhang-BSDE}) gives
\bea
&&\mbb{E}\Bigl[\sup_{t\in[0,T]}|\Del Y_t|^2+\int_0^T (|\Del Z_t^0|^2+|\Del Z_t|^2)dt\Bigr]\nn \\
&&\qquad\quad \leq C\mbb{E}\Bigl[|\Del X_T|^2+\int_0^T |\Del X_t|^2dt\Bigr]
+\zeta C\mbb{E}\Bigl[|\Del x_T|^2+\int_0^T (|\Del x_t|^2+|\Del y_t|^2) dt\Bigr]~.\nn
\eea
Combining with $(\ref{eq-pw-1})$ and choosing $\zeta>0$ small, we obtain
\bea
\mbb{E}\Bigl[\sup_{t\in[0,T]}|\Del Y_t|^2+\int_0^T (|\Del Z_t^0|^2+|\Del Z_t|^2)dt\Bigr]
\leq \zeta C\mbb{E}\Bigl[|\Del x_T|^2+\int_0^T (|\Del x_t|^2+|\Del y_t|^2)dt\Bigr]~.
\label{eq-pw-2}
\eea
By the similar procedures, we also have
\bea
\mbb{E}\Bigl[\sup_{t\in[0,T]}|\Del X_t|^2\Bigr]\leq \zeta C\mbb{E}\Bigl[|\Del x_T|^2+\int_0^T (|\Del x_t|^2+|\Del y_t|^2)dt\Bigr]~.
\label{eq-pw-3}
\eea
From $(\ref{eq-pw-2})$ and $(\ref{eq-pw-3})$, we obtain with $C=C(L,\ol{\Lambda}, \gamma_c, T,\zeta)$
\bea
&&\mbb{E}\Bigl[\sup_{t\in[0,T]}|\Del X_t|^2+\sup_{t\in[0,T]}|\Del Y_t|^2+\int_0^T (|\Del Z_t^0|^2+|\Del Z_t|^2)dt\Bigr]\nn \\
&&\leq \zeta C\mbb{E}\Bigl[\sup_{t\in[0,T]}|\Del x_t|^2+\sup_{t\in[0,T]}|\Del y_t|^2+\int_0^T (|\Del z_t^0|^2+|\Del z_t|^2)dt\Bigr]\nn,
\eea
where $C$ is decreasing as $\zeta$ goes to zero.
Thus there exists $\zeta>0$, being independent of the size of $\vr$, that makes the map $(x,y,z^0,z)\mapsto (X,Y,Z^0,Z)$
strict contraction. Therefore the initial hypothesis holds true for $(\vr+\zeta)$, which establishes the existence.
The uniqueness follows from the next proposition.
\end{proof}
\end{theorem}

\begin{proposition}
\label{prop-stability}
Given two set of inputs $(\xi, c^0, c), (\xi\pp,c^{0\p},c\pp)$, coefficients $(b,\L), (b\pp,\L\pp)$
and the coefficient functions $(l,\sigma_0,\sigma, \ol{f},\ol{g}), (l\pp,\sigma_0\pp, \sigma\pp, \ol{f}\pp,\ol{g}\pp)$ satisfying  
Assumptions~\ref{MFG-a}, \ref{MFG-b} and \ref{MFG-c1}, 
let us denote the corresponding solutions to $(\ref{fbsde-single-p})$ by $(X,Y,Z^0,Z)$ and $(X\pp,Y\pp,Z^{0\p},Z\pp)$,
respectively. We also define the functionals $(B,F, G)$and $(B\pp,F\pp, G\pp)$ by $(\ref{BG-notation})$
with corresponding coefficients, respectively. Then, we have the following stability result:
\bea
&&\mbb{E}\Bigl[\sup_{t\in[0,T]}|\Del X_t|^2+\sup_{t\in[0,T]}|\Del Y_t|^2+\int_0^T (|\Del Z_t^0|^2+|\Del Z_t|^2)dt\Bigr]\nn \\
&&\qquad \leq C\mbb{E}\Bigl[|\Del \xi|^2+|\del G|^2+\int_0^T\Bigl(
|\del F(t)|^2+|\del B(t)|^2+|\del \sigma_0(t)|^2+|\del \sigma(t)|^2\Bigr)dt\Bigr]~, \nn
\eea
where $C$ is a constant depending only on $T$ as well as the Lipschitz constants of the system, and
\bea
&&\del B(t):=B(t,Y_t\pp,c^0_t,c_t)-B\pp(t,Y_t\pp,c_t^{0\p},c_t\pp), \nn \\
&&\del F(t):=F(t,X_t\pp,Y_t\pp,c_t^0,c_t)-F\pp(t,X_t\pp,Y_t\pp, c_t^{0\prime},c_t\pp), \nn \\
&&(\del \sigma_0, \del \sigma)(t):=\bigl(\sigma_0(t,c_t^0,c_t)-\sigma_0\pp(t,c_t^{0\p},c_t\pp), \sigma(t,c_t^0,c_t)-\sigma\pp(t,c_t^{0\p},c_t\pp)\bigr), \nn \\
&&\del G:=G(X_T\pp,c_T^0,c_T)-G\pp(X_T\pp,c_T^{0\p},c_T\pp)~, \nn
\eea
and $\Del \xi:=\xi-\xi\pp$, $\Del X_t:=X_t-X_t\pp$ and similarly for the other variables. 
\begin{proof}
Let us put  $\Del B(t):=B(t,Y_t,c_t^0,c_t)-B(t,Y_t\pp,c_t^0,c_t)$,  
$\Del F(t):=F(t,X_t,Y_t, c_t^0,c_t)-F(t,X_t\pp,Y_t\pp, c_t^0,c_t)$ and
$\Del G:=G(X_T,c_T^0,c_T)-G(X_T\pp,c_T^0,c_T)$.
We get by \Ito's formula to $\langle \Del X, \Del Y\rangle$,
\bea
&&\mbb{E}\bigl[\langle \Del X_T, \Del G+\del G\rangle\bigr]=\mbb{E}\Bigl[\langle \Del \xi,\Del Y_0\rangle+\int_0^T \Bigl(\langle 
\del F(t), \Del X_t\rangle
+\langle \del B(t), \Del Y_t\rangle \nn \\
&&\qquad+\langle \del \sigma_0(t), \Del Z_t^0\rangle+\langle \del \sigma(t), \Del Z_t\rangle
+\bigl(\langle \Del F(t),\Del X_t \rangle+\langle \Del B(t),\Del Y_t\rangle\bigr)\Bigr)dt\Bigr]~.\nn
\eea
Using $(\ref{peng-wu-condition})$, we obtain
\bea
&&\gamma \mbb{E}\Bigl[|\Del X_T|^2+\int_0^T |\Del X_t|^2 dt\Bigr]
\leq \mbb{E}\Bigl[\langle \Del \xi, \Del Y_0\rangle-\langle \Del X_T, \del G\rangle \nn \\
&&\qquad +\int_0^T\Bigl(\langle \del F(t),\Del X_t\rangle+\langle \del B(t),\Del Y_t\rangle+
\langle \del \sigma_0(t), \Del Z_t^0\rangle+\langle \del \sigma(t), \Del Z_t\rangle\Bigr)dt\Bigr]~.
\label{stability-1}
\eea
On the other hand, the standard estimates for Lipschitz SDEs and BSDEs give
\bea
\label{stability-y}
&&\mbb{E}\Bigl[\sup_{t\in[0,T]}|\Del Y_t|^2+\int_0^T (|\Del Z_t^0|^2+|\Del Z_t|^2)dt\Bigr] \nn \\
&&\qquad \leq C\mbb{E}\Bigl[|\del G|^2+\int_0^T |\del F(t)|^2 dt\Bigr]+C\mbb{E}\Bigl[|\Del X_T|^2+\int_0^T |\Del X_t|^2dt\Bigr]~,  \\
%%%%%%%%%%%%%
&&\mbb{E}\Bigl[\sup_{t\in[0,T]}|\Del X_t|^2\Bigr]
\leq C\mbb{E}\Bigl[|\Del \xi|^2+\int_0^T\bigl[|\del B(t)|^2+|\del \sigma_0(t)|^2+|\del \sigma(t)|^2\bigr]dt\Bigr]+C\mbb{E}\int_0^T |\Del Y_t|^2 dt~. \nn
\eea
Combining the above inequalities $(\ref{stability-1})$ and $(\ref{stability-y})$ gives
\bea
&&\mbb{E}\Bigl[\sup_{t\in[0,T]}|\Del X_t|^2+\sup_{t\in[0,T]}|\Del Y_t|^2+\int_0^T (|\Del Z_t^0|^2+|\Del Z_t|^2)dt\Bigr]\nn \\
&&\leq C\mbb{E}\Bigl[|\Del \xi|^2+|\del G|^2+\int_0^T\bigl[|\del F(t)|^2+|\del B(t)|^2+|\del \sigma_0(t)|^2+|\del \sigma(t)|^2\bigr]dt\Bigr]\nn \\
&&+C\mbb{E}\Bigl[\langle \Del \xi, \Del Y_0\rangle-\langle \Del X_T, \del G\rangle 
+\int_0^T\bigl[\langle \del F(t),\Del X_t\rangle+\langle \del B(t),\Del Y_t\rangle+
\langle \del \sigma_0(t), \Del Z_t^0\rangle+\langle \del \sigma(t), \Del Z_t\rangle\bigr]dt\Bigr]~.\nn
\eea
Now simple application of Young's inequality establishes the claim.
\end{proof}
\end{proposition}

\begin{corollary}
\label{L2-estimate}
Under Assumptions~\ref{MFG-a}, \ref{MFG-b} and \ref{MFG-c1}, the solution $(X,Y,Z^0,Z)$ to the FBSDE $(\ref{fbsde-single-p})$ satisfies the 
following estimate:
\bea
&&\mbb{E}\Bigl[\sup_{t\in[0,T]}|X_t|^2+\sup_{t\in[0,T]}|Y_t|^2+\int_0^T (|Z_t^0|^2+|Z_t|^2)dt\Bigr] 
\leq C\mbb{E}\Bigl[|\xi|^2+|\part_x\ol{g}(0,c_T^0,c_T)|^2 \nn \\
&&\hspace{40mm}+\int_0^T\Bigl(|\part_x\ol{f}(t,0,0,c_t^0,c_t)|^2+|l(t,0,c_t^0,c_t)|^2+
|(\sigma_0,\sigma)(t,c_t^0,c_t)|^2\Bigr)dt\Bigr]~,\nn
\eea
where $C$ is a constant depending only on $T, b$ and Lipschitz constants of the system.
\begin{proof}
By quick inspection of the proof for Proposition~\ref{prop-stability}, one sees that
we have used properties of the coefficient functions given in Assumptions~\ref{MFG-a}, \ref{MFG-b} and \ref{MFG-c1}
only for $(l,\sigma_0,\sigma,\ol{f},\ol{g})$.
In fact, we would have gotten the same conclusion in the proposition even if 
the coefficients $(l\pp,\sigma_0\pp, \sigma\pp, \ol{f}\pp,\ol{g}\pp)$ were chosen
arbitrarily as long as  $(X\pp, Y\pp,Z^{0\p},Z\pp)\in \mbb{S}^2\times \mbb{S}^2\times \mbb{H}^2\times \mbb{H}^2$ were well-defined.
In particular, by putting $\xi\pp$ and $(l\pp,\sigma_0\pp, \sigma\pp, \ol{f}\pp,\ol{g}\pp)$
all zero, we have a trivial solution $(X\pp,Y\pp,Z^{0\pp},Z\pp)=(0,0,0,0)$. The desired estimate now follows from 
Proposition~\ref{prop-stability}.
\end{proof}
\end{corollary}
%%%%%%%%%%%%%%%%%%%%%
\subsection{Application to securities with exogenously specified payoff}
%%%%%%%%%%%%%%%%%%%%%
If we consider the exchange markets of bonds and futures, or other financial derivatives with maturity $T$, 
those securities cease to exist at $T$ after paying exogenously specified amount of cash $c_T^0$. 
In this case, as we have mentioned in Remark~\ref{remark-3-1}, it is natural to consider with $b=0$ and
\bea
g(x,c^0)=\ol{g}(x,c^0):=-\langle c^0,x\rangle~,
\label{terminal-special}
\eea
since there is no reason to put penalty on the outstanding volume at $T$.
In this case, the terminal function $g$ in $(\ref{terminal-special})$ does not have the strict convexity.
Fortunately, even in this case, we can prove the unique existence as well as the stability result of the same form.

\begin{assumption} 
\label{MFG-c2}{~}\\
%(MFG-c2)\\
%{\rm (i)} The functions $\sigma_0$ and $\sigma$ are independent of the argument $\vp$. \\
%{\rm (ii)} For any $t\in[0,T]$,  any random variables $x,x\pp, c^0, c\in \mbb{L}^2(\calf;\mbb{R}^n)$ and any sub-$\sigma$-field $\calg\subset \calf$, 
%the function $l$ satisfies the monotone condition with some positive constant $\gamma^l>0$:
%\bea
%\mbb{E}\Bigl[\langle l(t,\mbb{E}[x|\calg],c^0,c)-l(t,\mbb{E}[x\pp|\calg],c^0,c),x-x\pp\rangle\Bigr]
%\geq \gamma^l\mbb{E}\bigl[\mbb{E}[x-x\pp|\calg]^2\bigr]~.\nn
%\eea
The same conditions as in Assumption~\ref{MFG-c1} except {\rm (iii)}, which is replaced by\\ 
${\rm (iii)}^\prime$ $\gamma:=\gamma^f-\frac{L_\vp^2}{4\gamma^l}$ is strictly positive and the terminal function $g$ is given by $(\ref{terminal-special})$ with $b=0$.
\end{assumption}

\begin{proposition}
\label{prop-general-T-s}
Under Assumptions~\ref{MFG-a}, \ref{MFG-b} and \ref{MFG-c2},  there exists a unique strong solution $(X,Y,Z^0,Z)\in \mbb{S}^2(\mbb{F}^1;\mbb{R}^n)
\times \mbb{S}^2(\mbb{F}^1;\mbb{R}^n)\times \mbb{H}^2(\mbb{F}^1;\mbb{R}^{n\times d^0})\times \mbb{H}^2(\mbb{F}^1;\mbb{R}^{n\times d})$
to the FBSDE $(\ref{fbsde-single-p})$. Moreover, the same form of stability and $\mbb{L}^2$ estimates given in 
Proposition~\ref{prop-stability} and Corollary~\ref{L2-estimate} hold.
\begin{proof}
Note that, in this case, the terminal condition for the BSDE is independent of $X_T$.
Thus, as in \cite[Theorem 2.3]{Peng-Wu}, we put $y_T^\vr=Y_T=-c_T^0$ in $(\ref{shifted-0})$ and $(\ref{shifted-1})$, respectively.
Using the fact that $\langle \Del X_T,\Del Y_T\rangle=0$, one can follow the same arguments to get the desired result.
The proof of the stability result can also be done in almost exactly the same way.
\end{proof}
\end{proposition}

Before closing the section, 
let us give a  simple example of $n$ bonds which have coupon streams represented by $n$-dimensional process $(c_t^0)_{t\in[0,T)}$
as well as the unit principal payment at the maturity $T$. We introduce two 
measurable functions $p:[0,T]\rightarrow \mbb{R}^n$ and  $q:[0,T]\times \mbb{R}^n\rightarrow \mbb{R}^n$
satisfying
\be
|p(t)|\leq C, \quad |q(t,c)|\leq C(1+|c|)
\label{eq-example}
\ee
with some constant $C>0$ for any $(t,c)\in [0,T]\times \mbb{R}^n$.
\begin{corollary}
Suppose that  $\sigma_0, \sigma$ are constant matrices with appropriate dimensions, 
$c_T^0=(1,\cdots,1)^\top$ the $n$-dimensional vector, the functions $p$ and $q$ as above,
and that the terminal cost is given by  $(\ref{terminal-special})$.
Moreover,  the functions  of running cost  $\ol{f}:[0,T]\times (\mbb{R}^n)^3\rightarrow \mbb{R}$
as well as the OTC order flows  $l:[0,T]\times \mbb{R}^n\rightarrow \mbb{R}$ are given by
\be
\begin{split}
\ol{f}(t,x,c^0,c)&:=-\langle c^0,x\rangle+\frac{\vr}{2}\bigl|x-q(t,c)\bigr|^2,  \\
l(t,\vp)&:=\zeta (\vp-p(t)),
\end{split} \nn
\ee
respectively, where  $\vr, \zeta$ are positive constants. Then, there exists a unique solution to the 
FBSDE $(\ref{fbsde-single-p})$ with corresponding coefficients.
\end{corollary}
In the above  example,  except the noise terms,  the OTC clients of each agent 
tend to sell the $i$th bond (and hence  increase the storage level of the agents) when $\vp_t^i\geq p^i(t)$.
The function $q(t,c)$ denotes some target level of the storage depending on the idiosyncratic information $c$.

%%%%%%%%%%%%%%%%%%%%%%%%%%%%%%%%
\section{Asymptotic Market Clearing} 
\label{sec-asymptotic}
%%%%%%%%%%%%%%%%%%%%%%%%%%%%%%%%
We are now ready to investigate if our FBSDE $(\ref{fbsde-single-p})$ actually provides an
approximate of the market price and if so, how accurate it is.
By Theorem~\ref{th-individual}, if we use  $(-\mbb{E}\bigl[Y_t|\ol{\calf}_t^0\bigr])_{t\in[0,T]}$
as the input $(\vp_t)_{t\in[0,T]}$, where $(Y_t)_{t\in[0,T]}$ is the unique solution to the FBSDE $(\ref{fbsde-single-p})$
with the convention $\xi=\xi^1$ and $c=c^1$, 
the optimal strategy of the individual agent is given by
\bea
\ha^i_{\rm{mf}}(t):=\ha(Y_t^i, -\mbb{E}[Y_t|\ol{\calf}_t^0])=-\ol{\L}(Y_t^i-\mbb{E}[Y_t|\ol{\calf}_t^0])
\label{mf-alpha}
\eea
where $(Y^i_t)_{t\in[0,T]}$ is the solution to $(\ref{agent-FBSDE})$ with $(\vp_t=-\mbb{E}\bigl[Y_t|\ol{\calf}_t^0\bigr])_{t\in[0,T]}$.
The next theorem shows that the market clearing condition in the large-$N$ limit $(\ref{large-N-clearing})$ holds.

\begin{theorem}
\label{th-clearing}
If the conditions for Theorem~\ref{th-short-T}, Theorem~\ref{th-general-T} or Proposition~\ref{prop-general-T-s} are satisfied then we have
\bea
\lim_{N\rightarrow \infty} \mbb{E}\int_0^T \Bigl|\frac{1}{N}\sum_{i=1}^N\ha^i_{\rm{mf}}(t)\Bigr|^2 dt=0~.\nn 
\eea
Moreover if there exists some constant $\Gamma$ such that $\sup_{t\in[0,T]}\mbb{E}\bigl[|Y_t|^q\bigr]^\frac{1}{q}\leq \Gamma< \infty$ for some $q>4$, 
then there exists  some constant $C$ independent of $N$ such that
\bea
\mbb{E}\int_0^T \Bigl|\frac{1}{N}\sum_{i=1}^N\ha^i_{\rm{mf}}(t)\Bigr|^2 dt\leq C\Gamma^2 \ep_N,
\label{Glivenko-Cantelli}
\eea
where $\ep_N:=N^{-2/\max(n,4)}\bigl(1+\log(N)\bold{1}_{\{n=4\}}\bigr)$.
\begin{proof}
Let us consider the following set of FBSDEs with $1\leq i\leq N$ on the filtered probability space $(\Omega,\calf,\mbb{P};\mbb{F})$
constructed in Section~\ref{sec-notation}.
\bea
d\ul{X}_t^i&=&\Bigl(-\ol{\L}(\ul{Y}_t^i-\mbb{E}[\ul{Y}_t^i|\ol{\calf}_t^0])+l(t,-\mbb{E}[\ul{Y}_t^i|\ol{\calf}_t^0],c_t^0,c_t^i)\Bigr)dt\nn \\
&&\qquad+\sigma_0(t,-\mbb{E}[\ul{Y}_t^i|\ol{\calf}_t^0],c_t^0,c_t^i)dW_t^0+\sigma(t,-\mbb{E}[\ul{Y}_t^i|\ol{\calf}_t^0],c_t^0,c_t^i)dW_t^i, \nn \\
d\ul{Y}_t^i&=&-\part_x \ol{f}(t,\ul{X}_t^i,-\mbb{E}[\ul{Y}_t^i|\ol{\calf}_t^0],c_t^0,c_t^i)dt+\ul{Z}_t^{i,0}dW_t^0+\ul{Z}_t^i dW_t^i, \nn 
\eea
with $\ul{X}_0^i=\xi^i$ and 
\be
\ul{Y}_T^i=\frac{b}{1-b}\mbb{E}[\part_x\ol{g}(\ul{X}_T^i,c_T^0,c_T^i)|\ol{\calf}_T^0]+\part_x \ol{g}(\ul{X}_T^i,c_T^0,c_T^i).\nn
\ee
Thanks to the existence of unique strong solution, Yamada-Watanabe Theorem for FBSDEs (see, \cite[Theorem 1.33]{Carmona-Delarue-2}), 
there exists some measurable function $\Phi$ such that for every $1\leq i\leq N$,
\bea
(\ul{X}^i_t,\ul{Y}^i_t)_{t\in[0,T]}=\Phi\Bigl((c_t^0)_{t\in[0,T]}, (W_t^0)_{t\in[0,T]}, \xi^i, (c_t^i)_{t\in[0,T]}, (W_t^i)_{t\in[0,T]}\Bigr)~. \nn
\eea
Hence, conditionally on $\ol{\calf}^0$, the set of proceses $(\ul{X}^i_t, \ul{Y}^i_t)_{t\in[0,T]}$ with $1\leq i\leq N$ are independently and 
identically distributed. In particular, we have $\mbb{P}$-a.s.
\bea
&&\mbb{E}[\ul{Y}_t^i|\ol{\calf}_t^0]=\mbb{E}[Y_t|\ol{\calf}_t^0],\quad \forall t\in[0,T], \nn \\
&&\mbb{E}[\part_x\ol{g}(\ul{X}_T^i,c_T^0,c_T^i)|\ol{\calf}_T^0]=\mbb{E}[\part_x\ol{g}(X_T,c_T^0,c_T)|\ol{\calf}_T^0]~.
\label{law-identity}
\eea
Note that, under the convention $\xi^1=\xi$ and $c^1=c$, we actually have $(\ul{X}^1,\ul{Y}^1)=(X,Y)$.
From  $(\ref{law-identity})$, we conclude that
$(X^i_t,Y^i_t,Z^{i,0}_t,Z^i_t)_{t\in[0,T]}=(\ul{X}^i_t,\ul{Y}^i_t,\ul{Z}^{i,0}_t,\ul{Z}_t^i)_{t\in[0,T]}$
in $\mbb{S}^2(\mbb{F}^i)\times \mbb{S}^2(\mbb{F}^i)\times \mbb{H}^2(\mbb{F}^i)\times \mbb{H}^2(\mbb{F}^i)$.
Therefore,
\bea
\frac{1}{N}\sum_{i=1}^N\ha^i_{\rm{mf}}(t)=-\ol{\L}\Bigl(\frac{1}{N}\sum_{i=1}^N  \ul{Y}_t^i-\mbb{E}[\ul{Y}_t^1|\ol{\calf}_t^0]\Bigr)~.
\label{alpha-m}
\eea

We can easily check that
\bea
\mbb{E}\Bigl[W_2\Bigl(\frac{1}{N}\sum_{i=1}^N \del_{\ul{Y}_t^i}, \call(\ul{Y}_t^1|\ol{\calf}_t^0)\Bigr)^2\Bigr|\ol{\calf}_t^0\Bigr]
\leq\frac{2}{N}\sum_{i=1}^N \mbb{E}\bigl[|\ul{Y}_t^i|^2|\ol{\calf}_t^0\bigr]+2\mbb{E}\bigl[|\ul{Y}_t^1|^2|\ol{\calf}_t^0\bigr]
=4\mbb{E}\bigl[|\ul{Y}_t^1|^2|\ol{\calf}_t^0\bigr]~.\nn
\eea
Since $(\ul{Y}_t^i)_{1\leq i\leq N}$ are $\ol{\calf}^0_t$-conditionally independently and identically distributed
and also $\ul{Y}^1\in \mbb{S}^2$,  the same arguments leading to $(2.14)$ in \cite{Carmona-Delarue-2} 
imply that the pointwise convergence holds:
\bea
\lim_{N\rightarrow \infty}\mbb{E}\Bigl[W_2\Bigl(\frac{1}{N}\sum_{i=1}^N \del_{\ul{Y}_t^i}, \call(\ul{Y}_t^1|\ol{\calf}_t^0)\Bigr)^2\Bigr]=0~.
\label{pointwise-conv}
\eea
We are now going to show that the set of functions, $(f_N)_{N\in \mbb{N}}$ defined by
\bea
[0,T]\ni t\mapsto f_N(t):=\mbb{E}\bigl[W_2\bigl(\ol{\mu}_t, \mu_t\bigr)^2\bigr]\in \mbb{R} \nn
\eea
with $\ol{\mu}_t:=\frac{1}{N}\sum_{i=1}^N\del_{\ul{Y}_t^i}$ and $\mu_t:=\call(\ul{Y}_t^1|\ol{\calf}_t^0)$ are precompact in the set $\calc([0,T]; \mbb{R})$ endowed with the topology of uniform convergence.
In fact, uniformly in $N$,
\bea
\sup_{t\in[0,T]}|f_N(t)|\leq 4\sup_{t\in[0,T]}\mbb{E}\bigl[|\ul{Y}_t^1|^2\bigr]\leq C<\infty
\label{fN-inequality}
\eea
where $C$ is given by the estimate in Corollary~\ref{L2-estimate}.
Moreover,  for any $0\leq t,s\leq T$,  Cauchy-Schwarz, $(\ref{fN-inequality})$ and the triangular inequalities give
\bea
&&|f_N(t)-f_N(s)|\leq \mbb{E}\Bigl[\Bigl(W_2(\ol{\mu}_t,\mu_t)+W_2(\ol{\mu}_s,\mu_s)\Bigr)^2\Bigr]^\frac{1}{2}\mbb{E}\Bigl[\Bigl(W_2(\ol{\mu}_t,\mu_t)-W_2(\ol{\mu}_s,\mu_s)\Bigr)^2\Bigr]^\frac{1}{2}\nn \\
&&\quad \leq C\mbb{E}\Bigl[\Bigl(W_2(\ol{\mu}_t,\mu_t)-W_2(\ol{\mu}_s,\mu_s)\Bigr)^2\Bigr]^\frac{1}{2} \leq C\mbb{E}\Bigl[ W_2(\ol{\mu}_t,\ol{\mu}_s)^2+W_2(\mu_t,\mu_s)^2\Bigr]^\frac{1}{2}~. \nn \\
&&\quad \leq C\mbb{E}\Bigl[ \frac{1}{N}\sum_{i=1}^N |\ul{Y}_t^i-\ul{Y}_s^i|^2+|\ul{Y}_t^1-\ul{Y}_s^1|^2\Bigr]^\frac{1}{2}\nn \\
&&\quad \leq C\mbb{E}\bigl[|\ul{Y}_t^1-\ul{Y}_s^1|^2\bigr]^\frac{1}{2}~,\nn
\eea
uniformly in $N$, where we have used the fact that $(\ul{Y}^i)_{i\geq 1}$ are conditionally i.i.d at the last inequality.
Since $(\ul{Y}_t^1)_{t\in[0,T]}$ is a continuous process, the above estimate tells that
$(f_N)_{N\in \mbb{N}}$ is equicontinuous, which is also uniformly equicontinuous since we are working on the finite interval.
Now, Arzela-Ascoli theorem implies the desired precompactness.  

Combining with the pointwise convergence $(\ref{pointwise-conv})$,  we thus conclude
\bea
\lim_{N\rightarrow \infty}\sup_{t\in[0,T]}\mbb{E}\Bigl[W_2\Bigl(\frac{1}{N}\sum_{i=1}^N \del_{\ul{Y}_t^i}, \call(\ul{Y}_t^1|\ol{\calf}_t^0)\Bigr)^2\Bigr]=0~.
\label{law-conv}
\eea
Next, we claim that the following inequality holds:
\bea
\Bigl|\frac{1}{N}\sum_{i=1}^N  \ul{Y}_t^i-\mbb{E}[\ul{Y}_t^1|\ol{\calf}_t^0]\Bigr|\leq W_1\Bigl(\frac{1}{N}\sum_{i=1}^N \del_{\ul{Y}_t^i}, 
\call(\ul{Y}_t^1|\ol{\calf}_t^0)
\Bigr)~. \nn
\eea
The estimate can be proved as follows.
For any $\mu, \nu\in\calp_1(\mbb{R}^n)$, Jensen's inequality implies
\bea
\Bigl|\int_{\mbb{R}^n} x\mu(dx)-\int_{\mbb{R}^n}y\nu(dy)\Bigr|=\Bigl|\int_{\mbb{R}^{n}\times \mbb{R}^n}(x-y)\pi(dx,dy)\Bigr|
\leq \int_{\mbb{R}^n\times \mbb{R}^n}|x-y|\pi(dx,dy)\nn
\eea
for any coupling $\pi\in \Pi_1(\mu,\nu)$ with marginals $\mu$ and $\nu$.
Taking infimum over $\pi\in \Pi_1(\mu,\nu)$, the definition of Wasserstein distance in  $(\ref{def-W})$ gives
\bea
\Bigl|\int_{\mbb{R}^n} x\mu(dx)-\int_{\mbb{R}^n}y\nu(dy)\Bigr|\leq W_1(\mu,\nu). \nn
\eea
Substituting $\frac{1}{N}\sum_{i=1}^N \del_{\ul{Y}_t^i}$ and $\call(\ul{Y}_t^1|\ol{\calf}_t^0)$ into $\mu$ and $\nu$ in the above relation, 
we obtain the desired inequality.
Eq. $(\ref{alpha-m})$ and the obvious relation $W_1(\mu, \nu)\leq W_2(\mu,\nu)$ for $\mu, \nu \in \calp_2(\mbb{R}^n)$ give 
\bea
\mbb{E}\int_0^T \Bigl|\frac{1}{N}\sum_{i=1}^N\ha^i_{\rm{mf}}(t)\Bigr|^2 dt\leq C
\sup_{t\in[0,T]}\mbb{E}\Bigl[W_2\Bigl(\frac{1}{N}\sum_{i=1}^N \del_{\ul{Y}_t^i}, \call(\ul{Y}_t^1|\ol{\calf}_t^0)\Bigr)^2\Bigr]~. 
\label{law-conv-2}
\eea
The first conclusion now follows from $(\ref{law-conv})$. The latter claims directly follows from the expression $(\ref{law-conv-2})$
and the (Fourth Step) in the proof of \cite[Theorem 2.12]{Carmona-Delarue-2}.
\end{proof}
\end{theorem}

Theorem~\ref{th-clearing} justifies our intuitive understanding and a
special type of FBSDEs $(\ref{fbsde-single-p})$ derived in Section~\ref{sec-intuitive} as a reasonable 
model to approximate the market clearing price. 
When there exists higher integrability, Glivenko-Cantelli convergence theorem in the Wasserstein distance
even provides a specific order $\ep_N$ of convergence in terms of the number of agents $N$ (Eq. $(\ref{Glivenko-Cantelli})$).
See the discussions in \cite[Theorem 5.8, Remark 5.9]{Carmona-Delarue-1} for more details.

\begin{remark}
\label{remark-clearing}
In \cite{Fujii-Takahashi-conv}, we have shown that it is indeed possible to construct the
price process $(\vp_t^{(N)})_{t\in[0,T]}$ satisfying the market clearing condition $(\ref{orig-clearing})$ 
even when $N$ is finite, 
once we impose appropriate monotone conditions and also relax the information structure so that every agent $i$ has the perfect knowledge
including idiosyncratic shocks to the other agents, $j\in\{1,\cdots, N\}, j\neq i$. We also need to assume
that every agents behave as a price taker.
Interestingly, we showed that $(\vp_t^{(N)})_{t\in[0,T]}$ strongly converges to the process $\bigl(-\mbb{E}[Y_t|\ol{\calf}_t^0]\bigr)_{t\in[0,T]}$
studied in the current paper in the large-$N$ limit.
Moreover, the idiosyncratic information to the other agents becomes irrelevant in the same limit,
since the market price becomes adapted to $\ol{\mbb{F}}^0$ representing the common information.
This observation further supports our claim that $\bigl(-\mbb{E}[Y_t|\ol{\calf}_t^0]\bigr)_{t\in[0,T]}$
gives a reasonable approximation of the market price when the number of agents $N$ is large enough.
These contents are closely related to the backward propagation of chaos.
See, in particular, \cite[Remark 4.3]{Fujii-Takahashi-conv} and references therein.
\end{remark}

\begin{remark}
Consider the situation treated in Proposition~\ref{prop-general-T-s}, for example, a market model of a Futures contract.
If the contract pays unit amount of the 
underlying asset per contract whose value is exogenously given by $c_T^0$, our mean-field limit model $(\ref{fbsde-single-p})$
gives $Y_T=-c_T^0$. This means that the modeled Futures price satisfies $\vp_T=-\mbb{E}[Y_T|\ol{\calf}_T^0]=c_T^0$,
which guarantees the convergence of the modeled price  to the value of the underlying asset at the maturity $T$.
This is a crucially important feature that any market model of this type of securities must satisfy.
\end{remark}

%%%%%%%%%%%%%%%%%%%%%%%%%%%%%%%%%%%%%%%%%
\section{Extension to Multiple Populations}
\label{sec-multiple-p}
%%%%%%%%%%%%%%%%%%%%%%%%%%%%%%%%%%%%%%%%
The main limitation of the last model is that there exists only one type of agents who share
the common cost functions as well as the coefficient functions for their state dynamics.
Interestingly, it is rather straightforward to extend the model to the situation with multiple populations, where the agents in each 
population share the same cost and coefficient functions
but they can be different  population by population. 
From the perspective of the practical applications,  this is a big advantage  since we can analyze, for example,  the interactions 
between the Sell-side and Buy-side institutions for financial applications, 
or consumers and producers for economic applications.
For general issues of mean field games as well as
mean field type control problems in the presence of multiple populations without common noise, see Fujii~\cite{Fujii-mfg}.
Although there exists a common noise in the current model, the conditional law only enters as a form of expectation.
Therefore, as long as the system of FBSDEs  is Lipschitz continuous, there exists a unique strong solution at least for small $T$.
For general $T$, although it is rather difficult to find  an appropriate set of assumptions, it is still possible for some simple cases.
In this section, our main task is to find an appropriate limit model that extends $(\ref{fbsde-single-p})$ for 
multiple populations and the sufficient conditions that make appropriate monotone conditions hold,
which guarantees the existence of unique solution.
\\

In the following, we shall treat $m$ populations indexed by $p\in\{1,\cdots, m\}$.
For each $p$, $N_p\geq 1$ agents are assumed to belong to the population.
We denote by $(p,i)$ the ith agent in the population $p$. First, let us enlarge the probability space constructed in Section~\ref{sec-notation}.
In addition to $(\ol{\Omega}^0,\ol{\calf}^0,\ol{\mbb{P}}^0;\ol{\mbb{F}}^0)$, we introduce
$(\ol{\Omega}^{p,i},\ol{\calf}^{p,i},\ol{\mbb{P}}^{p,i};\ol{\mbb{P}}^{p,i})$ with  $1\leq i\leq N_p$ and $1\leq p\leq m$,
each of which is generated by $(\xi^{p,i},\bg{W}^{p,i})$ 
with  $d$-dimensional Brownian motion $\bg{W}^{p,i}$ and a $\bg{W}^{p,i}$-independent
 $\mbb{R}^n$-valued square integrable random variable $\xi^{p,i}$.
For each $p$, $(\xi^{p,i})_{i=1}^{N_p}$ are assumed to have the common law.
We define $(\Omega^{p,i},\calf^{p,i},\mbb{P}^{p,i};\mbb{F}^{p,i})$ as the product of 
$(\ol{\Omega}^0,\ol{\calf}^0,\ol{\mbb{P}}^0;\ol{\mbb{F}}^0)$ and $(\ol{\Omega}^{p,i},\ol{\calf}^{p,i},\ol{\mbb{P}}^{p,i};\ol{\mbb{P}}^{p,i})$.
Finally $(\Omega,\calf,\mbb{P};\mbb{F})$ is defined as a product of all the spaces $(\ol{\Omega}^0,\ol{\calf}^0,\ol{\mbb{P}}^0;\ol{\mbb{F}}^0)$
and
$(\ol{\Omega}^{p,i},\ol{\calf}^{p,i},\ol{\mbb{P}}^{p,i};\ol{\mbb{F}}^{p,i})$,
$1\leq i\leq N_p, 1\leq p\leq m$, and $(\Omega^i, \calf^i,\mbb{P}^i;\mbb{F}^i)$ as a product of 
$(\ol{\Omega}^0,\ol{\calf}^0,\ol{\mbb{P}}^0;\ol{\mbb{F}}^0)$ and $(\ol{\Omega}^{p,i},\ol{\calf}^{p,i},\ol{\mbb{P}}^{p,i};\ol{\mbb{F}}^{p,i})$
with $1\leq p\leq m$.
Every probability space is assumed to be complete and every filtration is assumed to be 
complete and right-continuously augmented to satisfy the  usual conditions.
\\

As we have done in Section~\ref{sec-intuitive}, we first assume that the market price of $n$ securities is given exogenously by
$\vp_t\in \mbb{H}^2(\ol{\mbb{F}}^0;\mbb{R}^n)$ with $\vp_T\in \mbb{L}^2(\ol{\calf}_T^0;\mbb{R}^n)$. 
Under this setup, we consider the control problem for each $(p,i)$ agent 
defined by 
\be
\inf_{\bg{\a}^{p,i}\in \mbb{A}^{p,i}}J^{p,i}(\bg{\a}^{p,i})~,
\label{control-pi}
\ee
with 
\bea
J^{p,i}(\bg{\a}^{p,i}):=\mbb{E}\Bigl[\int_0^T f_p(t,X_t^{p,i},\a_t^{p,i},\vp_t,c_t^0,c_t^{p,i})dt+g_p(X_T^{p,i},\vp_T,c_T^0,c_T^{p,i})\Bigr]~, \nn
\eea
subject to the dynamic constraint:
\bea
dX_t^{p,i}=\Bigl(\a_t^{p,i}+l_p(t,\vp_t,c_t^0,c_t^{p,i})\Bigr)dt+\sigma_{p,0}(t,\vp_t,c_t^0,c_t^{p,i})dW_t^0+\sigma_p(t,\vp_t,c_t^0,c_t^{p,i})dW_t^{p,i} \nn
\eea
with $X_0^{p,i}=\xi^{p,i}$. As before we assume $(c_t^0)_{t\geq 0}\in \mbb{H}^2(\ol{\mbb{F}}^0;\mbb{R}^n)$ with $c_T^0\in \mbb{L}^2(\ol{\calf}_T^0;
\mbb{R}^n)$
and $(c_t^{p,i})_{t\geq 0}\in \mbb{H}^2(\ol{\mbb{F}}^{p,i};\mbb{R}^n)$ with $c_T^{p,i}\in \mbb{L}^2(\ol{\calf}_T^{p,i};\mbb{R}^n)$. 
In addition, within each population $p$,
the random sources $(c^{p,i}_t)_{t\geq 0}$ are assumed to have a common law $1\leq i\leq N_p$.
Admissible strategies $\mbb{A}^{p,i}$ is the space $\mbb{H}^2(\mbb{F}^{p,i};\mbb{R}^n)$.
The measurable functions $f_p:[0,T]\times (\mbb{R}^n)^5\rightarrow \mbb{R}$, $g_p:(\mbb{R}^n)^4\rightarrow \mbb{R}$, 
$\ol{f}_p:[0,T]\times (\mbb{R}^n)^4\rightarrow \mbb{R}$ and $\ol{g}_p:(\mbb{R}^n)^3\rightarrow \mbb{R}$ 
are given by
\bea
&&f_p(t,x,\a,\vp,c^0,c):=\langle \vp,\a\rangle+\frac{1}{2}\langle \a,\L_p \a\rangle+\ol{f}_p(t,x,\vp,c^0,c)~, \nn \\
&&g_p(x,\vp,c^0,c):=-b \langle \vp,x\rangle+\ol{g}_p(x,c^0,c)~.\nn
\eea

\begin{assumption} %{\rm{(MFG-A)}} 
\label{MFG-A}
We assume the following conditions uniformly in $p\in\{1,\cdots,m\}$.\\
{\rm (i)} $\L_p$ is a positive definite $n\times n$ symmetric matrix with $\ul{\lambda}I_{n\times n}\leq \L_p \leq \ol{\lambda}I_{n\times n}$
in the sense of 2nd-order form where $\ul{\lambda}$ and $\ol{\lambda}$ are some constants
satisfying $0< \ul{\lambda}\leq \ol{\lambda}$.  \\
{\rm (ii)} For any $(t,x,\vp,c^0,c)$, 
\bea
|\ol{f}_p(t,x,\vp,c^0,c)|+|\ol{g}_p(x,c^0,c)|\leq L(1+|x|^2+|\vp|^2+|c^0|^2+|c|^2)~.\nn
\eea
{\rm (iii)} $\ol{f}_p$ and $\ol{g}_p$ are continuously differentiable in $x$ and satisfy,  for any $(t,x,x\pp, \vp,c^0,c)$,
\be
|\part_x \ol{f}_p(t,x\pp,\vp,c^0,c)-\part_x\ol{f}_p(t,x,\vp,c^0,c)|+|\part_x \ol{g}_p(x\pp,c^0,c)-\part_x \ol{g}_p(x,c^0,c)|\leq L |x\pp-x|~, \nn
\ee 
and $|\part_x \ol{f}_p(t,x,\vp,c^0,c)|+|\part_x \ol{g}_p(x,c^0,c)|\leq L(1+|x|+|\vp|+|c^0|+|c|)$. \\
{\rm (iv)} The functions $\ol{f}_p$ and $\ol{g}_p$ are convex in $x$ in the sense that for any $(t,x,x\pp,\vp,c^0,c)$,
\bea
&&\ol{f}_p(t,x\pp,\vp,c^0,c)-\ol{f}_p(t,x,\vp,c^0,c)-\langle x\pp-x, \part_x \ol{f}_p(t,x,\vp,c^0,c)\rangle\geq \frac{\gamma^f}{2}|x\pp-x|^2~, \nn \\
&&\ol{g}_p(x\pp,c^0,c)-\ol{g}_p(x,c^0,c)-\langle x\pp-x,\part_x \ol{g}_p(x,c^0,c)\rangle \geq \frac{\gamma^g}{2}|x\pp-x|^2~, \nn
\eea
with some constants $\gamma^f,\gamma^g\geq 0$. \\
{\rm (v)} $l_p,\sigma_{p,0},\sigma_p$ are the measurable functions defined on $[0,T]\times (\mbb{R}^n)^3$
and are $\mbb{R}^n, \mbb{R}^{n\times d^0}$ and $\mbb{R}^{n\times d}$-valued, respectively.
Moreover they satisfy the linear growth condition:
\bea
|(l_p,\sigma_{p,0},\sigma_p)(t,\vp,c^0,c)|\leq L(1+|\vp|+|c^0|+|c|) \nn
\eea
for any $(t,\vp,c^0,c)$. \\
{\rm (vi)} $b \in [0,1)$ is a given constant.
\end{assumption}

Under Assumption~\ref{MFG-A}, Theorem~\ref{th-individual} guarantees that the control problem $(\ref{control-pi})$
for each agent $(p,i)$ is uniquely characterized by 
\bea
&&dX_t^{p,i}=\Bigl(\ha_p(Y_t^{p,i},\vp_t)+l_p(t,\vp_t,c_t^0,c_t^{p,i})\Bigr)dt+\sigma_{p,0}(t,\vp_t,c_t^0,c_t^{p,i})dW_t^0+\sigma_p(t,\vp_t,c_t^0,c_t^{p,i})dW_t^{p,i}, \nn \\
&&dY_t^{p,i}=-\part_x \ol{f}_p(t,X_t^{p,i},\vp_t,c_t^0,c_t^{p,i})dt+Z_t^{p,i,0}dW_t^0+Z_t^{p,i}dW_t^{p,i},
\label{fbsde-pi}
\eea
with $X_0^{p,i}=\xi^{p,i}$ and $Y_T^{p,i}=-b \vp_T+\part_x \ol{g}_p(X_T^{p,i},c_T^0,c_T^{p,i})$. We have defined $\ha_p(y,\vp):=-\ol{\L}_p(y+\vp)$ and $\ol{\L}_p:=(\L_p)^{-1}$ as before. There exists a unique strong solution $(X^{p,i}_t,Y^{p,i}_t,Z^{p,i,0}_t,Z^{p,i}_t)_{t\in[0,T]}\in \mbb{S}^2(\mbb{F}^{p,i};\mbb{R}^n)
\times\mbb{S}^2(\mbb{F}^{p,i};\mbb{R}^n)\times \mbb{H}^2(\mbb{F}^{p,i};\mbb{R}^{n\times d^0}) \times \mbb{H}^2(\mbb{F}^{p,i};\mbb{R}^{n\times d})$,
and the optimal trading strategy for the agent $(p,i)$ is given by
\bea
\ha_t^{p,i}=\ha_p(Y_t^{p,i},\vp_t)~, \forall t\in[0,T]. \nn
\eea

Let us check the market clearing condition under this setup. 
In order to balance the demand and supply of the securities at the exchange, 
we need to have
 $\sum_{p=1}^m \sum_{i=1}^{N_p}\ha(Y^{p,i}_t,\vp_t)=0$. This requires the market price to satisfy
\bea
\vp_t=-\Bigl(\sum_{p=1}^m n_p \ol{\L}_p\Bigr)^{-1}\sum_{p=1}^m n_p\ol{\L}_p\Bigl(\frac{1}{N_p}\sum_{i=1}^{N_p}Y_t^{p,i}\Bigr)~, \nn
\eea 
where $N=\sum_{p=1}^m N_p$ and $n_p:=N_p/N$. At the moment, this is inconsistent to the 
initial assumption that requires $(\vp_t)_{t\geq 0}$ to be $\ol{\mbb{F}}^0$-adapted. 
However, since for each $1\leq p\leq m$, $(Y^{p,i}_t)_{i=1}^{N_p}$ are $\ol{\calf}^0$-conditionally
independently and identically distributed, we may follow the same arguments used in Section~\ref{sec-intuitive}.
If we take $N\rightarrow \infty$ while keeping the relative size of populations $n_p$ constant,
we can expect to obtain
\bea
\vp_t=- \hat{\Xi}\sum_{p=1}^m \hat{\L}_p \mbb{E}[Y_t^{p,1}|\ol{\calf}_t^0]
\label{market-price-guess}
\eea
in the large population limit where
\be
\hat{\L}_p:=n_p\ol{\L}_p, \qquad \hat{\Xi}:=\Bigl(\sum_{p=1}^m \hat{\L}_p\Bigr)^{-1}~.\nn
\ee

\begin{remark}
When $\L_p=\L$ for every population $p$, one can easily check that $(\ref{market-price-guess})$ becomes
\be
\vp_t=-\sum_{p=1}^m n_p \mbb{E}[Y_t^{p,1}|\ol{\calf}_t^0]~.\nn
\ee 
Since $Y$ of the adjoint equation represents the marginal cost i.e.,  the first order derivative 
of the value function with respect to the state variable $x$, the above expression of $\vp$ implies that the market price 
may be given by the population-weighted average of the marginal benefit (-cost) across the entire populations.
\end{remark}
%%%%%%%%%%%%%%%%
\subsection{Limit problem with multiple populations}
%%%%%%%%%%%%%%%%
By the observation we have just made, we are motivated to study the following limit problem with $1\leq p\leq m$:
\bea
dX_t^p&=&\Bigl(\ha_p\bigl(Y_t^p, \vp(\mbb{E}[Y_t|\ol{\calf}_t^0])\bigr)+l_p\bigl(t,\vp(\mbb{E}[Y_t|\ol{\calf}_t^0]),c_t^0,c_t^p\bigr)\Bigr)dt\nn \\
&&+\sigma_{p,0}\bigl(t,\vp(\mbb{E}[Y_t|\ol{\calf}_t^0]),c_t^0,c_t^p\bigr)dW_t^0+\sigma_p\bigl(t,\vp(\mbb{E}[Y_t|\ol{\calf}_t^0]),c_t^0,c_t^p\bigr)dW_t^{p,1}, \nn \\
dY_t^p&=&-\part_x\ol{f}_p\bigl(t,X_t^p, \vp(\mbb{E}[Y_t|\ol{\calf}_t^0]),c_t^0,c_t^p\bigr)dt+Z_t^{p,0}dW_t^0+Z_t^{p}dW_t^{p,1}~, 
\label{fbsde-multiple-p}
\eea
with $X_0^p=\xi^p$ and
\bea
Y_T^p=\frac{b}{1-b} \hat{\Xi}\sum_{p=1}^m \hat{\L}_p\mbb{E}\bigl[\part_x\ol{g}_p(X_T^p,c_T^0,c_T^p)|\ol{\calf}_T^0\bigr]+
\part_x\ol{g}_p(X_T^p,c_T^0,c_T^p)~. \nn
\eea
We put as before $\xi^p:=\xi^{p,1}$ and $c^p:=c^{p,1}$ to lighten the notation. Here, 
\bea
\vp(\mbb{E}[Y_t|\ol{\calf}_t^0]):=- \hat{\Xi}\sum_{p=1}^m \hat{\L}_p \mbb{E}[Y_t^{p}|\ol{\calf}_t^0], \quad \ha_p(y,\vp):=-\ol{\L}_p(y+\vp)\nn
\eea
and hence $(\ref{fbsde-multiple-p})$ is actually an $m$-coupled system of FBSDEs of McKean-Vlasov type. 
One can derive the terminal condition from
\be
Y_T^p=-b \vp(\mbb{E}[Y_T|\ol{\calf}_T^0])+\part_x \ol{g}_p(X_T^{p},c_T^0,c_T^{p})~, 
\label{terminal-pre}
\ee
by summing over $1\leq p\leq m$ after taking conditional expectation given $\ol{\calf}_T^0$.
In the following, we use the notation
\be
(X_t,Y_t,Z_t^0,Z_t)_{t\in[0,T]}=\Bigl((X_t^p)_{p=1}^m, (Y_t^p)_{p=1}^m, (Z_t^{p,0})_{p=1}^m, (Z_t^{p})_{p=1}^m\Bigr)_{t\in[0,T]}~.
\label{vector-notation}
\ee

%%%%
\subsection{Solvability for small $T$}
For small $T$,  Lipschitz continuity suffices to guarantee the existence of a unique solution.
\begin{assumption}%{\rm{(MFG-B)}} \\
\label{MFG-B}
Uniformly in $p\in\{1,\cdots,m\}$, for any $(t,x,c^0,c)\in[0,T]\times (\mbb{R}^n)^3$ and any $\vp,\vp\pp \in \mbb{R}^n$, 
the coefficient functions $l_p,\sigma_{p,0},\sigma_p$ and $\ol{f}_p$ satisfy,  with $L$ and another  nonnegative constant $L_\vp$,
\be
\begin{split}
&|(l_p,\sigma_{p,0},\sigma_p)(t,\vp,c^0,c)-(l_p,\sigma_{p,0},\sigma_p)(t,\vp\pp,c^0,c)|\leq L|\vp-\vp\pp|,  \\
&|\part_x \ol{f}_p(t,x,\vp,c^0,c)-\part_x \ol{f}_p(t,x,\vp\pp,c^0,c)|\leq L_\vp|\vp-\vp\pp|~.\nn
\end{split}
\ee
\end{assumption}
The next theorem follows exactly in the same way as Theorem~\ref{th-short-T}.

\begin{theorem}
\label{th-short-T-mp}
Under Assumptions~\ref{MFG-A} and \ref{MFG-B},  there exists some constant $\tau>0$ which depends only on $(L,L_\vp,b, n_p, \L_p)$
such that for any $T\leq \tau$, there exists a unique strong solution $(X,Y,Z^0,Z)\in \mbb{S}^2\bigl(\mbb{F}^1;(\mbb{R}^n)^m\bigr)
\times \mbb{S}^2\bigl(\mbb{F}^1;(\mbb{R}^n)^m\bigr)\times \mbb{H}^2\bigl(\mbb{F}^1;(\mbb{R}^{n\times d^0})^m\bigr)\times 
\mbb{H}^2\bigl(\mbb{F}^1;(\mbb{R}^{n\times d})^m\bigr)$ to the FBSDE $(\ref{fbsde-multiple-p})$.
\end{theorem}

\begin{remark}
Note that the above system of FBSDEs becomes a linear-quadratic form by choosing $(l_p,\sigma_{p,0},\sigma_p,\ol{f}_p,\ol{g}_p)$
appropriately. In this case, the problem reduces to solving ordinary differential equations of Riccati type.
Therefore, the existence of a solution for a given $T$ can be tested, at least numerically, by checking the absence of a
``blow up" in its solution.
\end{remark}

%%%%%%%%%%%%%%%%%%%%%%
\subsection{Solvability for general $T$}
We now move on to the existence result of a unique solution for general $T$. 
It is very difficult to find general existence criteria for fully-coupled multi-dimensional FBSDEs.
At the moment, in order to apply well-known Peng-Wu's method, let us put the following simplifying assumptions.

\begin{assumption}\label{MFG-C1}{~}\\ 
%{\rm{(MFG-C1)}}\\
{\rm (i)} For every $1\leq p\leq m$, the functions $\sigma_{p,0}$ and $\sigma_p$ are independent of the argument $\vp$.\\
{\rm (ii)} $\L_p$=$\L$ and $n_p=1/m$ for every $p$. \\
{\rm (iii)} For any $t\in[0,T]$, any random variables $x^p,x^{p\prime}, c^0,c^p\in \mbb{L}^2(\calf;\mbb{R}^n)$
and any sub-$\sigma$-field $\calg\subset \calf$, the functions $(l_p)_{p=1}^m$ satisfy the monotone condition, with some positive constant 
$\gamma^l>0$,
\bea
\sum_{p=1}^m \mbb{E}\Bigl[\bigl\langle l_p\bigl(t,\mbb{E}[\ol{x}|\calg],c^0,c^0\bigr)-l_p\bigl(t,\mbb{E}[\ol{x}^\prime|\calg],c^0,c^p\bigr),
x^p-x^{p\prime}\bigr\rangle \Bigr]\geq m \gamma^l \bold{1}_{\{L_\vp>0\}} \mbb{E}\bigl[\mbb{E}[\ol{x}-\ol{x}^\prime|\calg]^2\bigr], \nn
\eea
where $\ol{x}:=\frac{1}{m}\sum_{p=1}^m x^p$ and similarly for $\ol{x}^\prime$.\\
{\rm (iv)} There exists a strictly positive constant $\gamma$ satisfying $0<\gamma\leq \Bigl(\gamma^f-\frac{L_\vp^2}{4\gamma^l}\Bigr)\wedge 
\gamma^g$.
Moreover, the functions $(\ol{g}_p)_{p=1}^m$ satisfy 
for any $x^p,x^{p\p}, c^0,c^p\in \mbb{L}^2(\calf;\mbb{R}^n)$ and any sub-$\sigma$-field $\calg\subset \calf$, 
\bea
&&\frac{b}{1-b}m^{-1}\mbb{E}\Bigl[\bigl\langle \sum_{p=1}^m \mbb{E}[\part_x \ol{g}_p(x^p,c^0,c^p)-
\part_x \ol{g}_p(x^{p\p}, c^0,c^p)|\calg], \sum_{p=1}^m (x^p-x^{p\p})\bigr\rangle\Bigr]\nn \\
&&\qquad\quad+\gamma^g\sum_{p=1}^m \mbb{E}[|x^p-x^{p\p}|^2] \geq \gamma \sum_{p=1}^m \mbb{E}[|x^p-x^{p\p}|^2]~.\nn
\eea
\end{assumption}
\begin{remark}
The conditions (iii) and (iv) in the above assumption are rather restrictive. 
The condition (iii) is satisfied, for example, if $l_p$ has a separable form $l_p=h(x)+h_p(c^0_t,c_t^p)$ with 
some function $h$, which is common to every population and strictly monotone. 
(iv) is also satisfied by requiring similar structure. Or,  since $\part_x\ol{g}_p$ is Lipschitz
continuous in $x$, the absolute value of the first term is bounded by
$\frac{b}{1-b}\max((L_p)_{p=1}^m)\sum_{p=1}^m \mbb{E}|x^p-x^{p\p}|^2$,
where the $L_p$ is the Lipschitz constant for $\part_x \ol{g}_p$. Thus the 
condition (iv) is satisfied if $b \max((L_p)_{p=1}^m)$ is sufficiently small.
\end{remark}

The next result is the counterpart of Theorem~\ref{th-general-T}.
\begin{theorem}
\label{th-general-T-m}
Under Assumptions~\ref{MFG-A}, \ref{MFG-B} and \ref{MFG-C1},  there exists a unique strong solution 
$(X,Y,Z^0,Z)\in \mbb{S}^2\bigl(\mbb{F}^1;(\mbb{R}^n)^m\bigr)
\times \mbb{S}^2\bigl(\mbb{F}^1;(\mbb{R}^n)^m\bigr)\times \mbb{H}^2\bigl(\mbb{F}^1;(\mbb{R}^{n\times d^0})^m\bigr)\times 
\mbb{H}^2\bigl(\mbb{F}^1;(\mbb{R}^{n\times d})^m\bigr)$ to the FBSDE $(\ref{fbsde-multiple-p})$.
Moreover, the same form of stability and $\mbb{L}^2$ estimates given in Proposition~\ref{prop-stability} and 
Corollary~\ref{L2-estimate} hold.
\begin{proof}
Under Assumption~\ref{MFG-C1}, $(\ref{fbsde-multiple-p})$ can be written as
\bea
&&dX_t^p=\Bigl\{-\ol{\Lambda}\Bigl(Y_t^p-\frac{1}{m}\sum_{p=1}^m \mbb{E}[Y_t^p|\ol{\calf}_t^0]\Bigr)+
l_p\Bigl(t,-\frac{1}{m}\sum_{p=1}^m \mbb{E}[Y_t^p|\ol{\calf}_t^0],c_t^0,c_t^p\Bigr)\Bigr\}dt\nn \\
&&\qquad\qquad+\sigma_{p,0}(t,c_t^0,c_t^p)dW_t^0+\sigma_p(t,c_t^0,c_t^p)dW_t^{p,1}, \nn \\
&&dY_t^p=-\part_x \ol{f}_p\Bigl(t,X_t^p,-\frac{1}{m}\sum_{p=1}^m \mbb{E}[Y_t^p|\ol{\calf}_t^0],c_t^0,c_t^p\Bigr)dt+
Z_t^{p,0}dW_t^0+Z_t^p dW_t^{p,1}, \nn
\eea
with $X_0^p=\xi^p$ and 
\be
Y_T^p=\frac{b}{1-b}\frac{1}{m}\sum_{p=1}^m\mbb{E}\bigl[\part_x\ol{g}_p(X_T^p,c_T^0,c_T^p)|\ol{\calf}_T^0\bigr]+
\part_x \ol{g}_p(X_T^p,c_T^0,c_T^p)~. \nn
\ee
For each $p$, let us define the functionals $B_p, F_p$ and $G_p$ for any $y^p,x^p,c^0,c^p \in \mbb{L}^2(\calf;\mbb{R}^n)$ with
$y:=(y^p)_{p=1}^m$, $x:=(x^p)_{p=1}^m$  and $c:=(c^p)_{p=1}^m$ by
\bea
&&B_p(t, y, c^0,c^p):=-\ol{\L}\Bigl(y^p-\frac{1}{m}\sum_{p=1}^m  \mbb{E}[y^p|\ol{\calf}_t^0]\Bigr)+l_p
\Bigl(t,-\frac{1}{m}\sum_{p=1}^m \mbb{E}[y^p|\ol{\calf}_t^0],c^0,c^p\Bigr)\nn \\
&&F_p(t,x^p,y,c^0,c^p):=-\part_x \ol{f}\Bigl(t,x^p,-\frac{1}{m}\sum_{p=1}^m \mbb{E}[y^p|\ol{\calf}_t^0],c^0,c^p\Bigr), \nn \\
&&G_p(x,c^0,c):=\frac{b}{1-b}\frac{1}{m}\sum_{p=1}^m \mbb{E}[\part_x\ol{g}_p(x^p,c^0,c^p)|\ol{\calf}_T^0]+\part_x\ol{g}_p(x^p,c^0,c^p) ~,\nn
\eea
and set $B(t,y,c^0,c):=(B_p(t,y,c^0,c^p))_{p=1}^m$, $F(t,x,y,c^0,c):=(F_p(t,x^p,y,c^0,c^p))_{p=1}^m$ and $G(x,c^0,c):=(G_p(x,c^0,c))_{p=1}^m$.
With $\Del y:=y-y\pp$ and $\Del x:=x-x\pp$, we have from Assumption~\ref{MFG-C1}(iii), 
\bea
&&\mbb{E}\Bigl[\langle B(t,y,c^0,c)-B(t,y\pp,c^0,c), \Del y\rangle\Bigr]
:=\sum_{p=1}^m \mbb{E}\Bigl[\langle B_p(t,y,c^0,c)-B_p(t,y\pp,c^0,c), \Del y^p\rangle\Bigr]  \nn \\
&&\leq -\sum_{p=1}^m \mbb{E}[\langle \Del y^p, \ol{\L}\Del y^p\rangle]+\frac{1}{m}
\mbb{E}\Bigl[\bigl\langle \sum_{p=1}^m  \mbb{E}[\Del y^p|\ol{\calf}_t^0],\ol{\L}\sum_{p=1}^m \Del y^p\bigr\rangle\Bigr]
-m\gamma^l \bold{1}_{\{L_\vp>0\}} \mbb{E}\Bigl[\Bigl(\frac{1}{m}\sum_{p=1}^m \mbb{E}[\Del y^p|\ol{\calf}_t^0]\Bigr)^2\Bigr] \nn \\
&&\leq -m\gamma^l \bold{1}_{\{L_\vp>0\}} \mbb{E}\Bigl[\Bigl(\frac{1}{m}\sum_{p=1}^m \mbb{E}[\Del y^p|\ol{\calf}_t^0]\Bigr)^2\Bigr].
\label{DelB-ineq}
\eea
There exists a orthogonal matrix $P$ such that $P^\top \ol{\L}P$ becomes diagonal.
Then working on the new basis $\hat{y}^p=P^\top \Del y^p$, $1\leq p\leq m$,  the last inequality of $(\ref{DelB-ineq})$
can be checked component by component $1\leq i\leq n$ by the fact 
$(\sum_{p=1}^m \hat{y}^p_i)^2\leq m \sum_{p=1}^m |\hat{y}^{p}_i|^2$.
Second, from Assumptions~\ref{MFG-A}(iv), \ref{MFG-B} and Cauchy-Schwarz inequality,  
\bea
&&\mbb{E}\Bigl[\langle F(t,x,y,c^0,c)-F(t,x\pp,y\pp,c^0,c),\Del x\rangle \Bigr] 
:=\sum_{p=1}^m \mbb{E}\Bigl[\langle F_p(t,x,y,c^0,c)-F_p(t,x\pp,y\pp,c^0,c),\Del x^p\rangle \Bigr] 
\nn \\
&&\quad\leq -\Bigl(\gamma^f-\frac{L_\vp^2}{4\gamma^l}\Bigr)\mbb{E}[|\Del x|^2]+m \gamma^l \bold{1}_{\{L_\vp>0\}}
\mbb{E}\Bigl[\Bigl(\frac{1}{m}\sum_{p=1}^m \mbb{E}[\Del y_t^p|\ol{\calf}_t^0]\Bigr)^2\Bigr]. 
\label{DelF-ineq}
\eea
Finally, from Assumptions~\ref{MFG-A} and \ref{MFG-C1}(iv), we immediately get
\bea
\mbb{E}\Bigl[\langle G(x,c^0,c)-G(x\pp,c^0,c), \Del x\rangle\Bigr]
:=\sum_{p=1}^m\mbb{E}\Bigl[\langle G_p(x,c^0,c)-G_p(x\pp,c^0,c), \Del x^p\rangle\Bigr]
\geq \gamma \mbb{E}[|\Del x|^2]~.\nn
\eea
Now we have established the monotone conditions corresponding to $(\ref{peng-wu-condition})$ for the current model.
We can now repeat the same procedures in the proof of Theorem~\ref{th-general-T} and Proposition~\ref{prop-stability}.
\end{proof}
\end{theorem}

Let us give the  results for the securities of maturity $T$ with exogenously 
specified payoff.
\begin{assumption}\label{MFG-C2}{~}\\
%{\rm{(MFG-C2)}}\\
%{\rm (i)} For every $1\leq p\leq m$, the functions $\sigma_{p,0}$ and $\sigma_p$ are independent of the argument $\vp$.\\
%{\rm (ii)} $\L_p$=$\L$ and $n_p=1/m$ for every $p$. \\
%{\rm (iii)} For any $t\in[0,T]$, any random variables $x^p,x^{p\prime}, c^0,c^p\in \mbb{L}^2(\calf;\mbb{R}^n)$
%and any sub-$\sigma$-field $\calg\subset \calf$, the functions $(l_p)_{p=1}^m$ satisfy with some positive constant $\gamma^l>0$,
%\bea
%\sum_{p=1}^m \mbb{E}\Bigl[\bigl\langle l_p\bigl(t,\mbb{E}[\ol{x}|\calg],c^0,c^0\bigr)-l_p\bigl(t,\mbb{E}[\ol{x}^\prime|\calg],c^0,c^p\bigr),
%x^p-x^{p\prime}\bigr\rangle \Bigr]\geq m \gamma^l \mbb{E}\bigl[\mbb{E}[\ol{x}-\ol{x}^\prime|\calg]^2\bigr], \nn
%\eea
%where $\ol{x}:=\frac{1}{m}\sum_{p=1}^m x^p$ and similarly for $\ol{x}^\prime$.\\
The same conditions as in Assumption~\ref{MFG-C1} except {\rm (iv)}, which is replaced by\\
${\rm (iv)}^\prime$ $\gamma:=\gamma^f-\frac{L_\vp^2}{4\gamma^l}$ is strictly positive. Moreover, $b=0$ and the terminal function $g_p$ is given by
\bea
g_p(x,c^0)=\ol{g}_p(x,c^0):=-\langle c^0, x\rangle
\eea 
for every $1\leq p\leq m$.
\end{assumption}

\begin{proposition}
\label{prop-general-T-m-s}
Under Assumptions~\ref{MFG-A}, \ref{MFG-B} and \ref{MFG-C2}, there exists a unique strong solution 
$(X,Y,Z^0,Z)\in \mbb{S}^2\bigl(\mbb{F}^1;(\mbb{R}^n)^m\bigr)
\times \mbb{S}^2\bigl(\mbb{F}^1;(\mbb{R}^n)^m\bigr)\times \mbb{H}^2\bigl(\mbb{F}^1;(\mbb{R}^{n\times d^0})^m\bigr)\times 
\mbb{H}^2\bigl(\mbb{F}^1;(\mbb{R}^{n\times d})^m\bigr)$ to the FBSDE $(\ref{fbsde-multiple-p})$.
Moreover, the same form of the stability and $\mbb{L}^2$ estimates given in Proposition~\ref{prop-stability}
and Corollary~\ref{L2-estimate} holds.
\begin{proof}
Using the inequalities $(\ref{DelB-ineq})$ and $(\ref{DelF-ineq})$ with $\sum_{p=1}^m \langle \Del X_T^p,\Del Y_T^p\rangle=0$,
we can follow the same arguments in the proof of Proposition~\ref{prop-general-T-s}.
\end{proof}
\end{proposition}

%%%%%%%%%%%%%%%%%%%%%%%%%%%%%%%%%%%%%%%%%%%%%%%%%%%%%%%%
\subsection{Asymptotic market clearing for multi-population model}
%%%%%%%%%%%%%%%%%%%%%%%%%%%%%%%%%%%%%%%%%%%%%%%%%%%%%%%%
At the last part of this section, we investigate the asymptotic market clearing in the presence of multiple populations.
As in Section~\ref{sec-asymptotic}, we define $(\vp_t)_{t\in[0,T]}$ using the solution to the system of the mean-field
FBSDEs:
\bea
\vp_t=\vp(\mbb{E}[Y_t|\ol{\calf}_t^0]):=-\hat{\Xi}\sum_{p=1}^m \hat{\L}_p\mbb{E}\bigl[Y_t^p|\ol{\calf}_t^0\bigr] \nn
\eea
where $(Y_t^p)_{p=1}^m$ is the solution of $(\ref{fbsde-multiple-p})$. 
In order to test the accuracy of the above $(\vp_t)_{t\in[0,T]}$ as a market clearing price, 
we solve the individual agent problem $(\ref{control-pi})$ with this  $\vp$ as an input.
The corresponding individual problem $(\ref{control-pi})$ for the agent $(p,i)$ is given by the unique strong solution 
$(X^{p,i},Y^{p,i},Z^{p,i,0},Z^{p,i})$ of $(\ref{fbsde-pi})$. The optimal strategy for the agent $(p,i)$ is then 
given by
\bea
\ha_{\rm{mf}}^{p,i}(t):=-\ol{\L}_p\Bigl(Y_t^{p,i}-\hat{\Xi}\sum_{q=1}^m \hat{\L}_q\mbb{E}\bigl[Y_t^q|\ol{\calf}_t^0\bigr]\Bigr)~, \forall t\in[0,T]~.\nn
\eea

\begin{theorem}
\label{th-clearing-m}
If the conditions for Theorem~\ref{th-short-T-mp}, Theorem~\ref{th-general-T-m} or Proposition~\ref{prop-general-T-m-s} are satisfied then we have
\bea
\lim_{N\rightarrow \infty} \mbb{E}\int_0^T \Bigl|\frac{1}{N}\sum_{p=1}^m \sum_{i=1}^{N_p}\ha^{p,i}_{\rm{mf}}(t)\Bigr|^2 dt=0~,\nn 
\eea
where $N:=\sum_{p=1}^m N_p$ and the limit is taken while keeping $(n_p:=N_p/N)_{1\leq p\leq m}$ constant.
Moreover if there exists some constant $\Gamma$ such that $\sup_{t\in[0,T]}\mbb{E}\bigl[|Y_t|^q\bigr]^\frac{1}{q}\leq \Gamma< \infty$ for some $q>4$, 
then there exists  some constant $C$ independent of $N$ such that
\bea
\mbb{E}\int_0^T \Bigl|\frac{1}{N}\sum_{p=1}^m\sum_{i=1}^{N_p}\ha^{p,i}_{\rm{mf}}(t)\Bigr|^2 dt
\leq C\Gamma^2 \ep_N,\nn 
\eea
where 
$ \ep_N:=N^{-2/\max(n,4)}\bigl(1+\log(N)\bold{1}_{\{n=4\}}\bigr)$.
\begin{proof}
By the definition of $\ha^{p,i}_{\rm{mf}}$, we have
\bea
\frac{1}{N}\sum_{p=1}^m \sum_{i=1}^{N_p}\ha^{p,i}_{\rm{mf}}(t)&=&
-\frac{1}{N}\sum_{p=1}^m \sum_{i=1}^{N_p}\ol{\L}_p\Bigl(Y_t^{p,i}-\hat{\Xi}\sum_{q=1}^m \hat{\L}_q \mbb{E}[Y_t^q|\ol{\calf}_t^0]\Bigr)\nn \\
&=&-\sum_{p=1}^m \hat{\L}_p\Bigl(\frac{1}{N_p}\sum_{i=1}^{N_p}Y_t^{p,i}-\mbb{E}[Y_t^p|\ol{\calf}_t^0]\Bigr)~.
\label{clearing-mp}
\eea
On the other hand, we have for each $1\leq p\leq m$, $1\leq i\leq N_p$, 
\bea
dX_t^{p,i}&=&\Bigl(\ha_p\bigl(Y_t^{p,i},\vp(\mbb{E}[Y_t|\ol{\calf}_t^0])\bigr)+l_p\bigl(t,\vp(\mbb{E}[Y_t|\ol{\calf}_t^0]),c_t^0,
c_t^{p,i}\bigr)\Bigr)dt\nn \\
&&+\sigma_{p,0}\bigl(t,\vp(\mbb{E}[Y_t|\ol{\calf}_t^0]),c_t^0,c_t^{p,i}\bigr)dW_t^0+
\sigma_p\bigl(t,\vp(\mbb{E}[Y_t|\ol{\calf}_t^0]),c_t^0,c_t^{p,i}\bigr)dW_t^{p,i}, \nn \\
dY_t^{p,i}&=&-\part_x\ol{f}_p\bigl(t,X_t^{p,i},\vp(\mbb{E}[Y_t|\ol{\calf}_t^0]),c_t^0,c_t^{p,i}\bigr)dt+Z_t^{p,i,0}dW_t^0+
Z_t^{p,i}dW_t^{p,i}~, \nn
\eea
with $X_0^{p,i}=\xi^{p,i}$,
\bea
Y_T^{p,i}=-b \vp(\mbb{E}[Y_T|\ol{\calf}_T^0])+\part_x\ol{g}_p(X_T^{p,i},c_T^0,c_T^{p,i})~. \nn
\eea
By the unique strong solvability,  Yamada-Watanabe theorem implies that there exists some function $\Phi_p$ for each $1\leq p\leq m$ such that
for every $1\leq i\leq N_p$,
\bea
(Y_t^{p,i})_{t\in[0,T]}
=\Phi_p\Bigl(c^0, (W_t^0)_{t\in[0,T]}, (\mbb{E}[Y_t^q|\ol{\calf}_t^0]_{t\in[0,T]})_{1\leq q\leq m}, \xi^{p,i}, (c_t^{p,i})_{t\in[0,T]}, (W^{p,i}_t)_{t\in[0,T]}\Bigr).\nn
\eea
Hence $(Y_t^{p,i})_{t\in[0,T], 1\leq i\leq N_p}$ are independently and identically distributed conditionally on $\ol{\calf}^0$.
In particular, we have $\mbb{E}[Y_t^{p,i}|\ol{\calf}_t^0]=\mbb{E}[Y_t^{p,1}|\ol{\calf}_t^0]$.

We now compare $(X^{p,1}_t,Y^{p,1}_t, Z^{p,1,0}_t, Z^{p,1}_t)_{t\in[0,T]}$ with 
$(X^p_t,Y^p_t,Z^{p,0}_t,Z^p_t)_{t\in[0,T]}$ by treating $\vp(\mbb{E}[Y_t|\ol{\calf}_t^0])$ as external inputs.
Note that the terminal condition of the latter satisfies the relation $(\ref{terminal-pre})$.
Then the standard stability result of the Lipschitz FBSDEs implies
$(Y^{p,1}_t)_{t\in[0,T]}=(Y^p_t)_{t\in[0,T]}$ in $\mbb{S}^2(\mbb{F}^{p,1};\mbb{R}^n)$. 
As a result we have obtained $\mbb{E}[Y_t^p|\ol{\calf}_t^0]=\mbb{E}[Y_t^{p,1}|\ol{\calf}_t^0]$.
Using the expression $(\ref{clearing-mp})$, we obtain
\bea
\frac{1}{N}\sum_{p=1}^m \sum_{i=1}^{N_p}\ha^{p,i}_{\rm{mf}}(t)&=&
-\sum_{p=1}^m \hat{\L}_p\Bigl(\frac{1}{N_p}\sum_{i=1}^{N_p}Y_t^{p,i}-\mbb{E}[Y_t^{p,1}|\ol{\calf}_t^0]\Bigr)~. \nn
\eea
We can now repeat the last part of the proof for Theorem~\ref{th-clearing}.
\end{proof}
\end{theorem}

%%%%%%%%%%%%%%%%%%%%%%%%
\section{Extensions and Concluding Remarks}
\label{sec-conclude}
%%%%%%%%%%%%%%%%%%%%%%%%

In this work, we have studied endogenous formation of market clearing price using a 
stylized model of the securities exchange. We have derived a special type of FBSDE
of McKean-Vlasov type with common noise whose solution provides an approximate of the 
equilibrium price. 
In addition to the existence of strong unique solution to the FBSDE,
we have proved that the modeled price asymptotically clear the market 
in the large $N$-limit. We also gave the order of convergence $\ep_N$ when the solution of the FBSDE possesses 
higher order of integrability.
In the following, let us list up of a further extension of our technique and some interesting topics 
for future projects:\\\\
%%%%%%%%%%%%%%%%%%%%
\bull{\bf{Dependence on the conditional law of the state}}:
For applications to energy and commodity markets, or economic models with producers and consumers, 
one may want to study the cost functions $(\ol{f}, \ol{g})$ depending on the empirical distribution of the state $X$ of the agents
such as
${\ol{f}}\Bigl(t,X_t^i , \frac{1}{N}\sum_{j=1}^N \del_{X_t^j}, \vp_t,c^0_t,c^i_t\Bigr)$.
Under the setup with conditional independence,  the cost function for the limit problem is 
naturally given by
${\ol{f}}\Bigl(t,X_t,\call(X_t|\ol{\calf}_t^0), \vp_t,c^0_t,c_t\Bigr)$.
Even in this case, the resultant FBSDE $(\ref{fbsde-single-p})$ is solvable, at least for small $T$, 
if $(\part_x \ol{f}, \part_x\ol{g})$ are Lipschitz continuous in the measure argument with respect to $W_2$-distance.
Under the stronger assumption guaranteeing the monotone conditions $(\ref{peng-wu-condition})$,
one can even achieve the existence of unique solution in general $T$.  As long as the common noise 
is solely from the filtration $\ol{\mbb{F}}^0$ generated by $\bg{W}^0$, we can avoid 
subtleties regarding the admissibility (so-called $H$-hypothesis). See 
\cite[Remark 2.10]{Carmona-Delarue-2} as a useful summary for this issue.
\\\\
%%%%%%%%%%%%%%%%%%%
\bull{\bf{Explicit solution}}: 
%%%%%%%%%%%%%%%%%%%%%%
If we chose $\ol{f},\ol{g}$ as quadratic functions and $l,\sigma_0,\sigma$ as affine functions,
we obtain a linear-quadratic mean field game with common noise. In this case,
an explicit solution may be available where the coefficients functions are 
given as the solutions to differential equations 
of Riccati type.
\\\\
%%%%%%%%%%%%%%%%%%%%%%%%%%%%%%
\bull{\bf{Property of market price process}}: It seems interesting 
to study the properties of the market clearing price theoretically and numerically.
For example, if $n=d^0$ the equivalent martingale measure (EMM) can be uniquely determined.
Based on the payoff distribution $c^0$ and the cost functions of the agents $(\ol{f},\ol{g})$,
one may study how the market price process under the EMM behaves,
for example, the relation between the skew of its implied volatility and the risk-averseness of the agents.
\\\\
%%%%%%%%%%%%%%%%%%%%%%%%%%%%%%%%%%%%%
\bull{\bf{Market clearing equilibrium with a major agent}}:
The problem of equilibrium price formation in the presence of a major agent is an important problem.
In our recent work Fujii \& Takahashi (2021)~\cite{Fujii-Takahashi-Major},
we provide the extension of the current model in the presence of a major financial firm
who has a non-negligible market share even in the large-population limit of the minor financial firms.

\subsubsection*{Acknowledgments.}
We thank two anonymous referees for valuable comments and feedbacks.
%%%%%%%%%%%%%%%%%%%%%%%%%%%%%%%%%%%%%%%%%%%%%%%%%%%%%%%%%%%%%%%%%%%%%

%%%%%%%%%%%%%%%%%%%%%%%%%%%%%%%%%%%%%%%%%%%%%%%%%%%%%%%%%%%%%%%%%%%%

\end{document}